\documentclass[iop]{emulateapj}

\usepackage{graphicx}
\usepackage{amsmath}
\usepackage[]{natbib}
\usepackage[usenames]{color} 
\usepackage{ulem}
\newcommand{\etal}{{et al.~}}

\newcommand\ie{{\it i.e.~}}

\newcommand{\kms}{\mbox{$\>{\rm km\, s^{-1}}$}}

\newcommand{\kpc}{\mbox{$\>{\rm kpc}$}} 
\newcommand{\pc}{\mbox{$\>{\rm pc}$}} 
\newcommand{\Gyr}{\mbox{$\>{\rm Gyr}$}}
\newcommand{\Myr}{\mbox{$\>{\rm Myr}$}}
\newcommand{\yr}{\mbox{$\>{\rm yr}$}}
\newcommand{\Msun}{\>{\rm M_{\odot}}}

\newcommand{\sig}[1]{\mbox{$\sigma_{#1}$}}

\newcommand{\feh}{\mbox{$\rm [Fe/H]$}}
\newcommand{\al}{\mbox{$\rm \alpha$}}
\newcommand{\ofe}{\mbox{$\rm [O/Fe]$}}
\newcommand{\alfe}{\mbox{$\rm [\alpha/Fe]$}}

\def\apjs{{ApJS}}

\shorttitle{Chemical dichotomy as an outcome of clumps}

\shortauthors{Clarke \etal}
\begin{document}

\title{The imprint of clump formation at high redshift. I. A disc $\alpha$-abundance dichotomy}

\author{Adam J.~Clarke\altaffilmark{1}}
\author{Victor P.~Debattista\altaffilmark{1}}
\author{David L. Nidever\altaffilmark{2,3}}
\author{Sarah R.~Loebman\altaffilmark{4,5}}
\author{Raymond C.~Simons\altaffilmark{6}}
\author{Susan Kassin\altaffilmark{6}}
\author{Min Du\altaffilmark{7}}
\author{Melissa Ness\altaffilmark{8,9}}
\author{Deanne B. Fisher\altaffilmark{10}}
\author{Thomas R.~Quinn\altaffilmark{11}}
\author{James Wadsley\altaffilmark{12}}
\author{Ken C. Freeman\altaffilmark{13}}
\author{Cristina C. Popescu\altaffilmark{1,14}}

\altaffiltext{1}{Jeremiah Horrocks Institute, University of Central
  Lancashire, Preston, PR1 2HE, UK; {\tt ajclarke90@gmail.com (AJC),
  vpdebattista@gmail.com (VPD), CPopescu@uclan.ac.uk (CCP)}}
\altaffiltext{2}{Department of Physics, Montana State University,
  P.O. Box 173840, Bozeman, MT 59717-3840}
\altaffiltext{3}{National Optical Astronomy Observatory, 950 North
  Cherry Ave, Tucson, AZ 85719, USA}
\altaffiltext{4}{{Department of Physics, University of California, Davis,
                 1 Shields Ave, Davis, CA 95616, USA}}
\altaffiltext{5}{Hubble Fellow}
\altaffiltext{6}{Johns Hopkins University, 3400 North Charles St.,
  Baltimore, MD 21218, USA}
\altaffiltext{7}{Kavli Institute of Astronomy and Astrophysics, Peking
  University, Beijing, 100871, China}
\altaffiltext{8}{Department of Astronomy, Columbia University, Pupin 
Physics Laboratories, New York, NY 10027, USA}
\altaffiltext{9}{Center for Computational Astrophysics, Flatiron Institute, 
162 Fifth Avenue, New York, NY 10010, USA}
\altaffiltext{10}{Centre for Astrophysics and Supercomputing, Swinburne
  University of Technology, P.O. Box 218, Hawthorn, VIC 3122,
  Australia}
\altaffiltext{11}{Astronomy Department, University of Washington, Box
  351580, Seattle, WA 98195, USA}
\altaffiltext{12}{Department of Physics and Astronomy, McMaster
  University, 1280 Main Street West, Hamilton, ON L8S 4M1, Canada}
\altaffiltext{13}{Research School of Astronomy and Astrophysics, Mount Stromlo Observatory, Cotter Road, Weston Creek ACT 2611, Australia}
\altaffiltext{14}{The Astronomical Institute of the Romanian Academy, Str. Cutitul de Argint 5, Bucharest, Romania}

\begin{abstract}
  The disc structure of the Milky Way is marked by a chemical
  dichotomy, with high-\al\ and low-\al\ abundance sequences,
  traditionally identified with the geometric thick and thin discs.
  This identification is aided by the old ages of the high-\al\ stars,
  and lower average ages of the low-\al\ ones.  Recent large scale
  surveys such as APOGEE have provided a wealth of data on this
  chemical structure, including showing that an identification of
  chemical and geometric thick discs is not exact, but the origin of
  the chemical dichotomy has remained unclear.  Here we demonstrate
  that a dichotomy arises naturally if the early gas-rich disc
  fragments, leading to some fraction of the star formation occuring
  in clumps of the type observed in high-redshift galaxies. These
  clumps have high star formation rate density.  They, therefore,
  enrich rapidly, moving from the low-\al\ to the high-\al\ sequence,
  while more distributed star formation produces the low-\al\
  sequence.  We demonstrate that this model produces a
  chemically-defined thick disc that has many of the properties of the
  Milky Way's thick disc.  Because clump formation is common in high
  redshift galaxies, we predict that chemical bimodalities are common
  in massive galaxies.
\end{abstract}


 \section{Introduction}
\label{s:intro.tex}

The Milky Way (MW) is important to our understanding of the structure
and evolution of galactic discs.  The geometical thick disc was
originally discovered in the MW by decomposing the vertical star count
profile into two components, the thin and thick discs
\citep{Yoshii1982, Gilmore1983}.  Spectroscopic studies of Solar
neighborhood stars found that the thick disc is composed of mainly
older ($\sim$12--7 Gyr) stars that are enhanced in their $\alpha$
abundances, while thin disc stars are younger ($<$7 Gyr) and have
near-Solar $\alpha$ abundances \citep[e.g.,][]{Fuhrmann1998,
Haywood2013}. \citet{Bovy2012b} studied the spatial distribution of
mono-abundance populations which showed smooth variations of
scale-height with abundance and suggested that there was no distinct
separation of ``geometrical'' discs \citep{Bovy2012a}, nor, perhaps,
in chemistry. Even so, the ``chemical'' separation of high and
low-$\alpha$ populations persists in recent high-resolution abundance
surveys such as APOGEE \citep[e.g.,][]{Anders2014, Nidever2014,
Hayden2015} and unbiased Solar neighborhood samples
\citep{Adibekyan2012}.

It is now believed that the chemically and geometrically
defined thick discs are different entities.  The chemically defined
thick disc has a shorter scale-length than the geometric thick disc
\citep{Robin1996, Ojha2001, Juric2008, Bensby2011, Bovy2012b, Martig2016}, 
although there is still some uncertainty about the scale-lengths
\citep{Robin2014}.  Old, \al-enhanced stars are concentrated to the 
inner MW \citet{Minchev2015}.  Stars in the chemical thick disc are
generally old, whereas geometric thick disc stars can have median ages
as young as $5\Gyr$ in the outer disc \citep{Martig2016}.

Several explanations have been proposed for the $\alpha$-bimodality.
The ``two-infall'' model \citep{Chiappini1997, Chiappini2009} suggests
that the early MW had a high star formation rate (short gas
consumption timescale) that changed dramatically around 8 Gyr when
there was a drop in the overall MW star formation rate and a large
infall of pristine gas that diluated the overall metallicity of the
Milky Way and gave rise to the thin disc, low-$\alpha$ population.
Since this model requires a particular event to occur (the large
infall of pristine gas at a particular time), it suggests that the
bimodality might not be a generic feature of all galaxies.

On the other hand, a ``superposition'' model (which is
similar to the scenario proposed in
\citealt{Schoenrich2009a}) postulates that the $\alpha$-bimodality is
a generic consequence of the chemical and dynamical processes in a
spiral disc galaxy.  The chemical evolution in $\alpha$ abundance is
quite rapid at early times but slows down at later times when most of
the stars are formed. In addition, a radial gradient in outflow rate
exists in the disc due to the decrease in the energy required to
unbind the gas.  The large outflow in the outer disc inhibits the gas
from attaining a high metallicity while in the inner galaxy chemical
evolution is able to advance fairly unimpeded.  Finally, the dynamical
process of radial migration then moves stars away from their birth
radii and thereby smoothes out the separate chemical evolution tracks.
The $\alpha$-bimodality is created by the fact that the chemical
evolution proceeds most quickly as SNIa's turn on and evolution is
moved from high-\al\ to low-\al\ thus producing fewer stars in the
$\alpha$ ``valley''.  Producing chemical bimodality via these
processes requires fine-tuning; indeed simulations which include these
processes have failed to produce any chemical bimodality
\citep[e.g.][]{Loebman2011, Minchev2013}, which instead
only produce a thick band in the ([$\alpha$/Fe], [Fe/H]) space
representing the overall chemical evolution of the galaxy with small
variations with radius.  \citet{Nidever2014} presents a detailed
discussion and example figures of the chemical evolution of these two
scenarios.

Nevertheless, in recent years some simulations have begun to find
hints of $\alpha$-bimodalities.  The EAGLE simulations have found
$\alpha$-bimodalities produced by an early centralized starbust
(important for the inner galaxy) and disc contraction (important for
the outer galaxy) that produce chemical bimodalities somewhat similar
to the MW in one of six simulated galaxies \citep{Grand2017}.  Others
have found that a late infall of a gas-rich galaxy can produce some
metal-poor, $\alpha$-poor stars that give rise to a bimodality
\citep{Snaith2016, Mackereth2018}. However, these bimodalities are
weaker than the one observed in the MW and the age distributions are
not always consistent with the MW, suggesting they are not the main
mechanism at work in the MW.

One option that has not been previously considered is the influence of
star forming ``clumps''.  Clumps are commonly found in {\it Hubble
Space Telescope} ({\it HST}) images of high redshift galaxies
\citep[e.g.][]{Elmegreen2005, Ravindranath2006, Elmegreen2007, 
ForsterSchreiber2011, Genzel2011, Guo2012, Guo2015}. Such clumps were
first identified by eye in {\it HST} images
\citep[e.g.][]{Cowie1995, vandenBergh1996} and are common in MW
progenitors at high redshift \citep[present in $\sim 55\%$ of such
galaxies at a redshift of 3,][]{Guo2015}.  They appear $\sim 1 \kpc$
in size when observed at low resolution, but at higher resolution
appear to have sizes $100\pc$ to $500\pc$ \citep{Livermore2012,
Livermore2015, Fisher2017, Cava2018}.  Clumps can contribute up to
$20\%$ of the integrated star-formation rate of a galaxy, but can be
less massive \citep[up to $7\%$ of the total mass,][]{Wuyts2012}. It
is proposed that they form a bulge in less than a half a Gyr
\citep{Dekel2009}.  They have also been proposed for the origin of the
geometrical thick disc \citep{Bournaud2009}.

In this paper, we demonstrate that discs evolving with significant
clump formation produce a bimodality in the [$\alpha$/Fe]-[Fe/H] space
by virtue of the very rapid star formation which occurs in clumps.  We
show that the resulting high-$\alpha$ track in chemical space is
comprised of old stars and suggest that this might be the process
forming the $\alpha$-bimodality in the MW.  The layout of this paper
is as follows.  In Section \ref{s:simulation} we describe the
simulation we use in this paper.  Section
\ref{s:discussion} then presents the results of the simulation.  We
present our conclusions in Section \ref{s:conclusion}.

 \section{Simulation}
\label{s:simulation}

We set up a spherical Navarro-Frenk-White \citep{Navarro1997} halo
with an equilibrium spherical gas distribution containing 10\%
of the total mass and following the same density distribution. The
starting halo is set up as described in \citet{Roskar2008}: it has a
mass within the virial radius ($r_{200} \simeq 200$ kpc) of $10^{12}
\Msun$. A temperature gradient ensures an initial gas pressure
equilibrium for an adiabatic equation of state. Gas velocities are
initialized to give a spin parameter of $\lambda = 0.065$
\citep{Bullock2001, Maccio2007}, with specific angular momentum $j
\propto R$, where $R$ is the cylindrical radius. Both the gas corona
and the dark matter halo are comprised of $10^6$ particles; gas
particles initially have masses $1.4 \times 10^5 \Msun$ and force
softening 50 pc.  Dark matter particles instead come in two mass
flavours ($10^6 \Msun$ and $3.5 \times 10^6 \Msun$ inside and outside
200 kpc, respectively), with a softening of 100 pc.  The stars in the
simulation form self-consistently out of cooling gas, inheriting the
softening and chemistry of the parent gas particle.

We evolve the simulation for 10 Gyr with {\sc gasoline}
\citep{Wadsley2004, Wadsley2017}, the smooth particle hydrodynamics (SPH) 
extension of the $N$-body tree-code {\sc pkdgrav} \citep{Stadel2001}.
Gas cooling includes the metal-line cooling implementation of
\citet{Shen2010}; in order to prevent the cooling from dropping below
the resolution of our simulation, we set a pressure floor on gas
particles $p_{floor} = 3 G\epsilon^2\rho^2$, where $G$ is Newton's
gravitational constant, $\epsilon$ is the softening length and $\rho$
is the gas particle's density \citep{Agertz+09}.
Gas cools and settles into a disc; once the density exceeds $1
\mathrm{cm}^{-3}$ and the temperature drops below 15,000~K, star
formation and supernova feedback cycles are initiated as described in
\citet{Stinson2006}.  A consequence of the metal-cooling is that gas
can reach lower temperatures where it is prone to fragmentation and
clump formation.  Supernovae feedback couples 10\% of the $10^{51}$
erg per supernova to the interstellar medium as thermal energy.  We
include the effects of turbulent diffusion \citep{Shen2010} allowing
the gas to mix, reducing the scatter in the age-metallicity relation
\citep{Pilkington2012}.  
We use a base time step of $\Delta t = 10$ Myr with timesteps refined
such that $\delta t = \Delta t/2^n < \eta\sqrt{\epsilon/a_g}$, where
$a_g$ is the acceleration at a particle's position and the refinement
parameter $\eta = 0.175$. We set the opening angle of the tree-code
gravity calculation to $\theta = 0.7$. In addition the time step of
gas particles satisfies the condition $\delta t_\mathrm{gas} =
\eta_\mathrm{courant} h/[(1+\alpha)c + \beta \mu_\mathrm{max}]$, where
$\eta_\mathrm{courant} = 0.4$, $h$ is the SPH smoothing length set
over the nearest 32 particles, $\alpha$ and $\beta$ are respectively
the linear and quadratic viscosity coefficients and $\mu_\mathrm{max}$
is described in \citet{Wadsley2004}.
Star particles represent single stellar populations with a
Miller-Scalo initial mass function.  SNII and SNIa yields of Oxygen
and Iron are taken from \citet{Raiteri1996}.  As in
\citet{Raiteri1996}, Padova stellar lifetimes are used to determine
SNII rates, and SNIa rates are determined from those same lifetimes in
a binary evolution model.

\subsection{Global properties of the final system}

\begin{figure*}[!h]
        \hskip -.3in \epsscale{1} \plotone{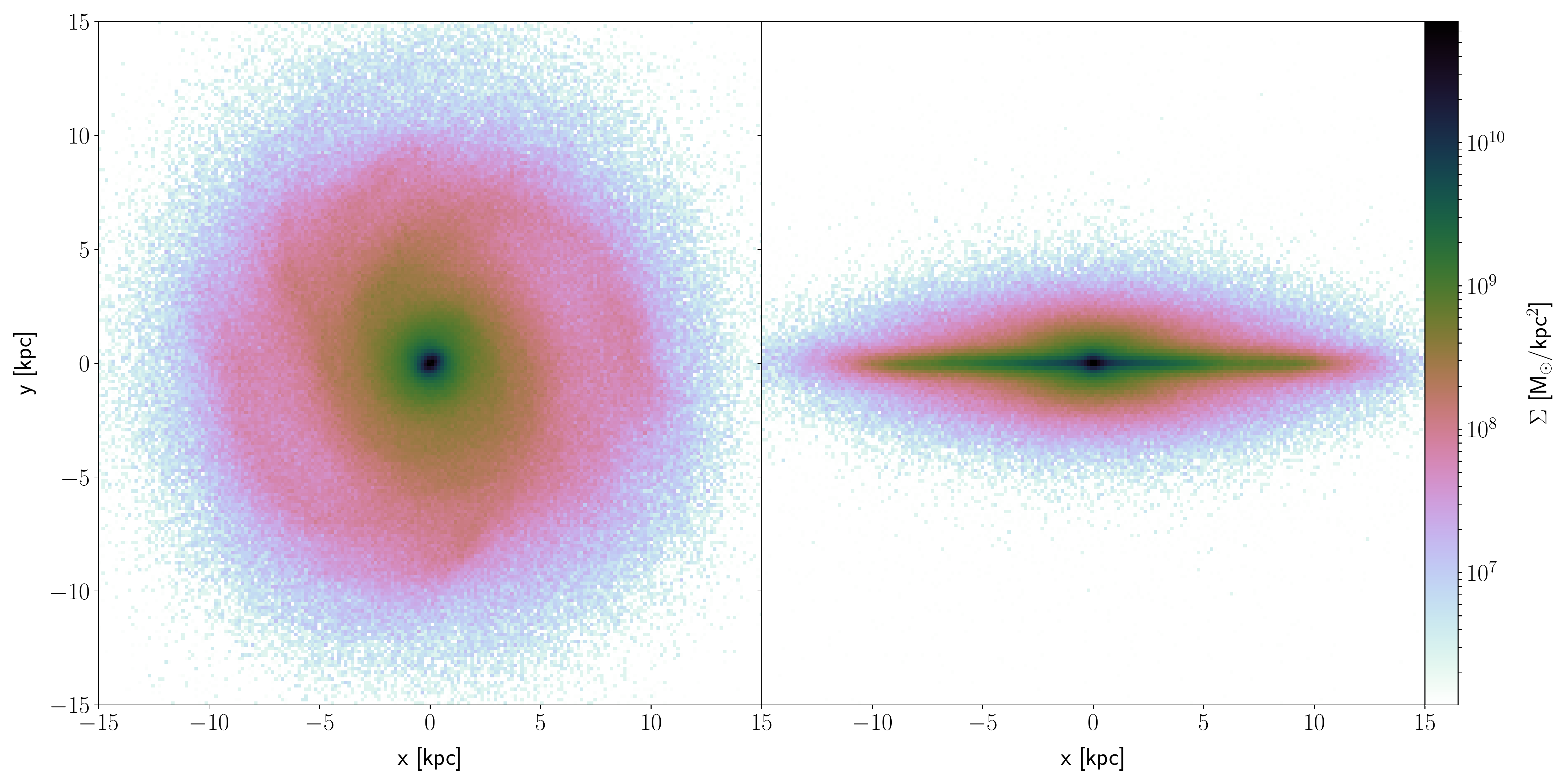}
        \caption{Stellar density of the final model viewed face-on
        (left) and edge-on (right).} \label{f:faceonedgeon}
\end{figure*}

\begin{figure}
\centerline{
\includegraphics[angle=0.,width=\hsize]{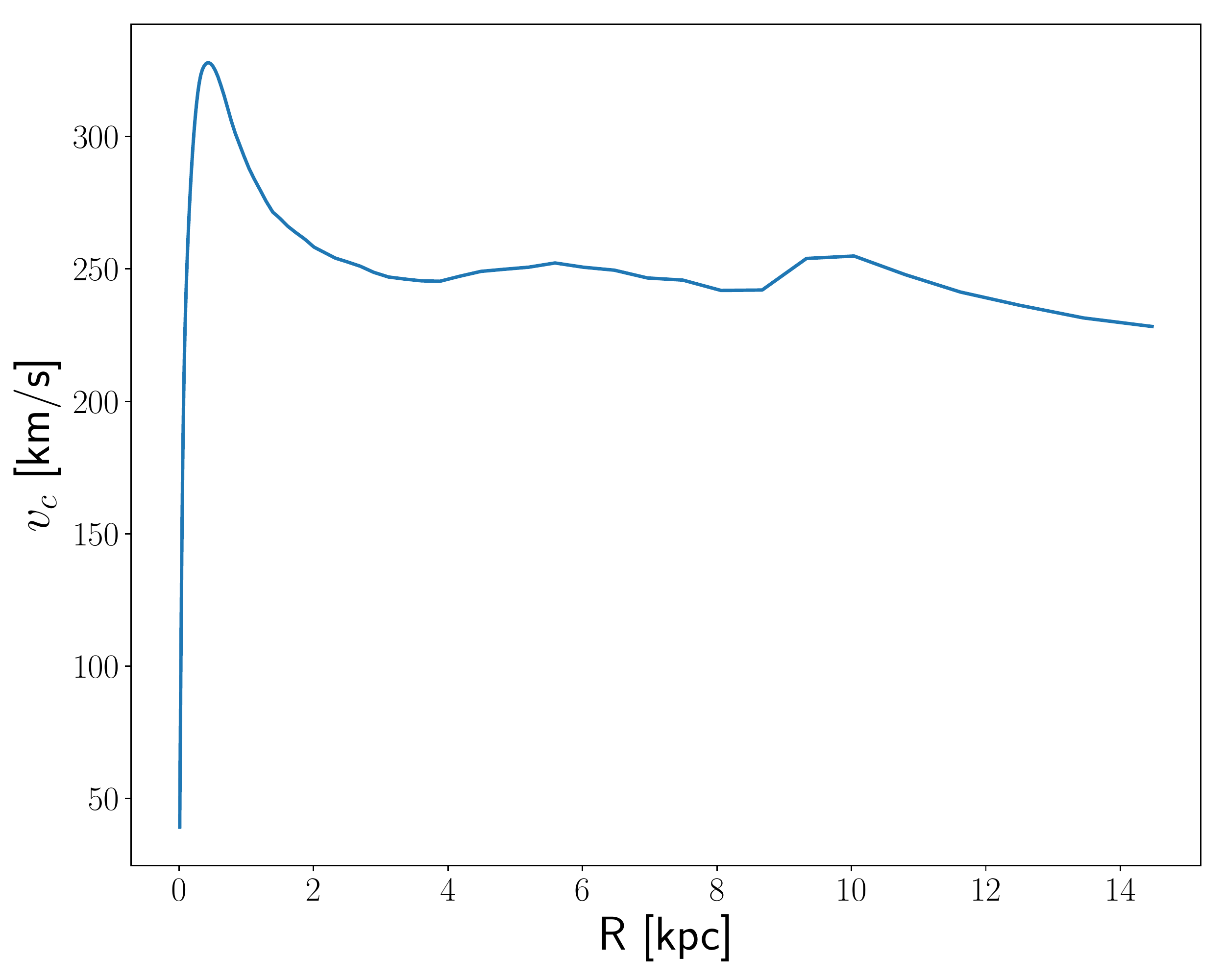}
}
\centerline{
\includegraphics[angle=0.,width=\hsize]{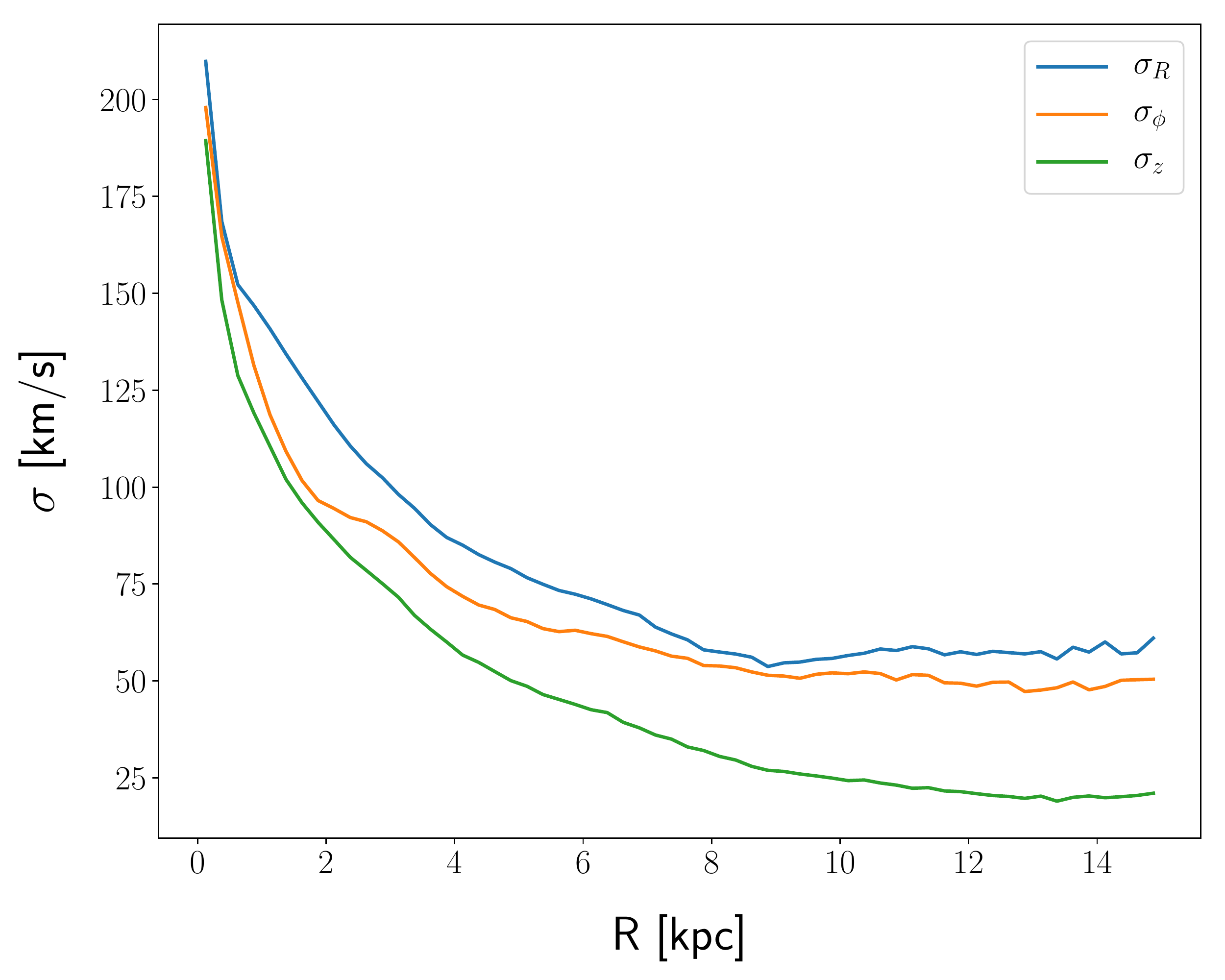}
}
\caption{Top: Rotation curve of the final model.  Bottom: Velocity 
dispersion profiles of all stars in different directions of the final
model.}
\label{f:kinematics}
\end{figure}

Figure \ref{f:faceonedgeon} presents the face-on and edge-on views of
the model at the end of the simulation, showing it to be a reasonable
example of a disc galaxy.  Figure \ref{f:kinematics} presents the
final rotation curve and velocity dispersions of the system.  Whilst
our simulation is not designed to match any specific galaxy, the
rotation curve of the final system has a circular velocity at $8 \kpc$
of $242~\kms$, making it comparable to the Milky Way.  Its structural
and kinematic properties are somewhat different from the Milky Way's,
for instance the model is unbarred, but is otherwise similar enough
that it can be used to understand the origin of trends observed in the
Milky Way.

 \section{Results}
\label{s:discussion}

\subsection{Chemical bimodality}

\begin{figure*}[!h]
	\hskip -.3in \epsscale{1}
	\plotone{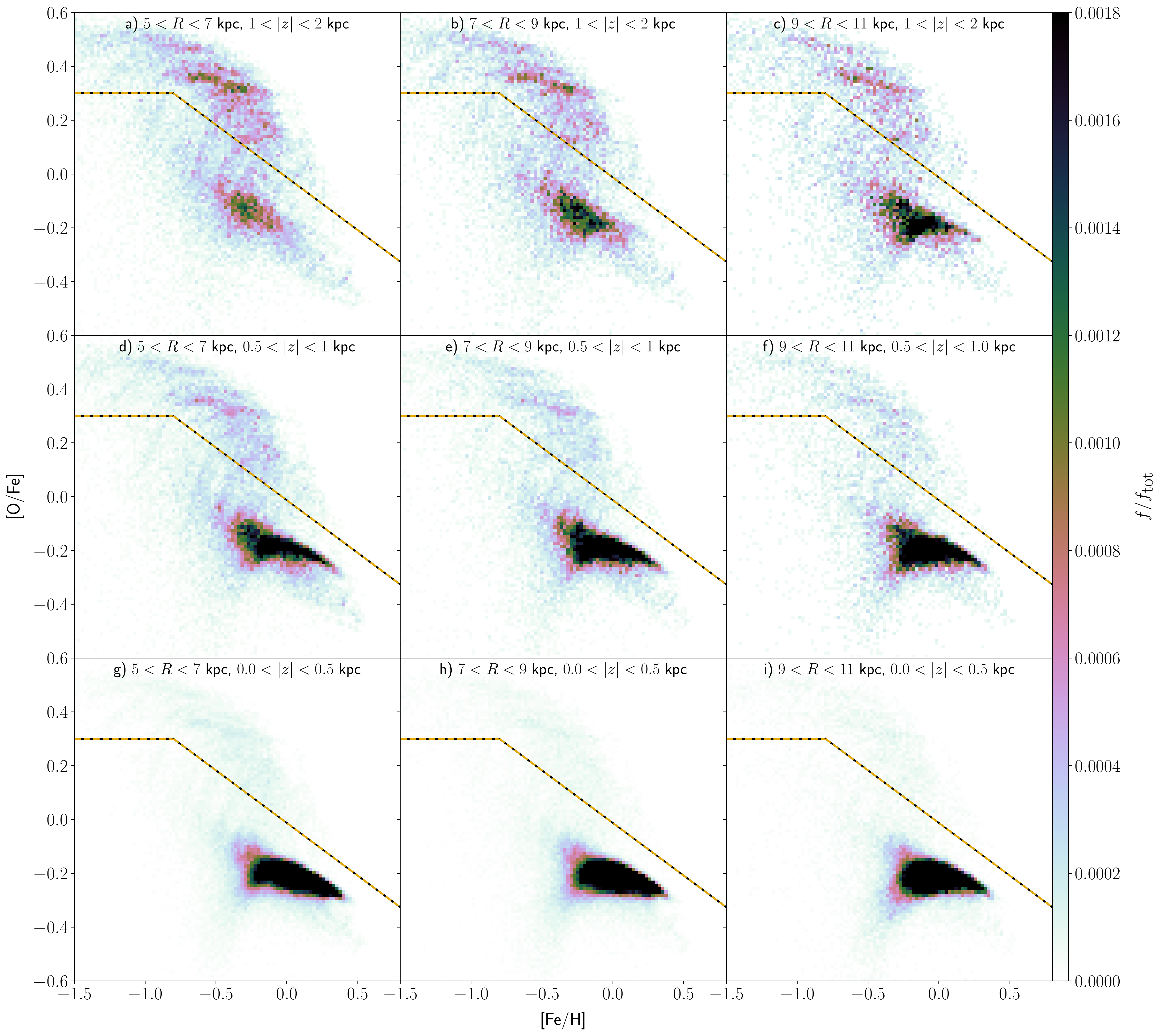}
	\caption{[O/Fe]-[Fe/H] distributions of the simulation for
	various radial and height cuts, as labelled in each panel. The
	dashed yellow-black line, defined by Eqn. \ref{e:sequences},
	separates the the high-\al\ and low-\al\ sequences.  Each
	panel is normalised to a total weight of unity.}
	\label{f:radialcuts}
\end{figure*}

\begin{figure}[!h]
        \hskip -.3in \epsscale{1}
        \plotone{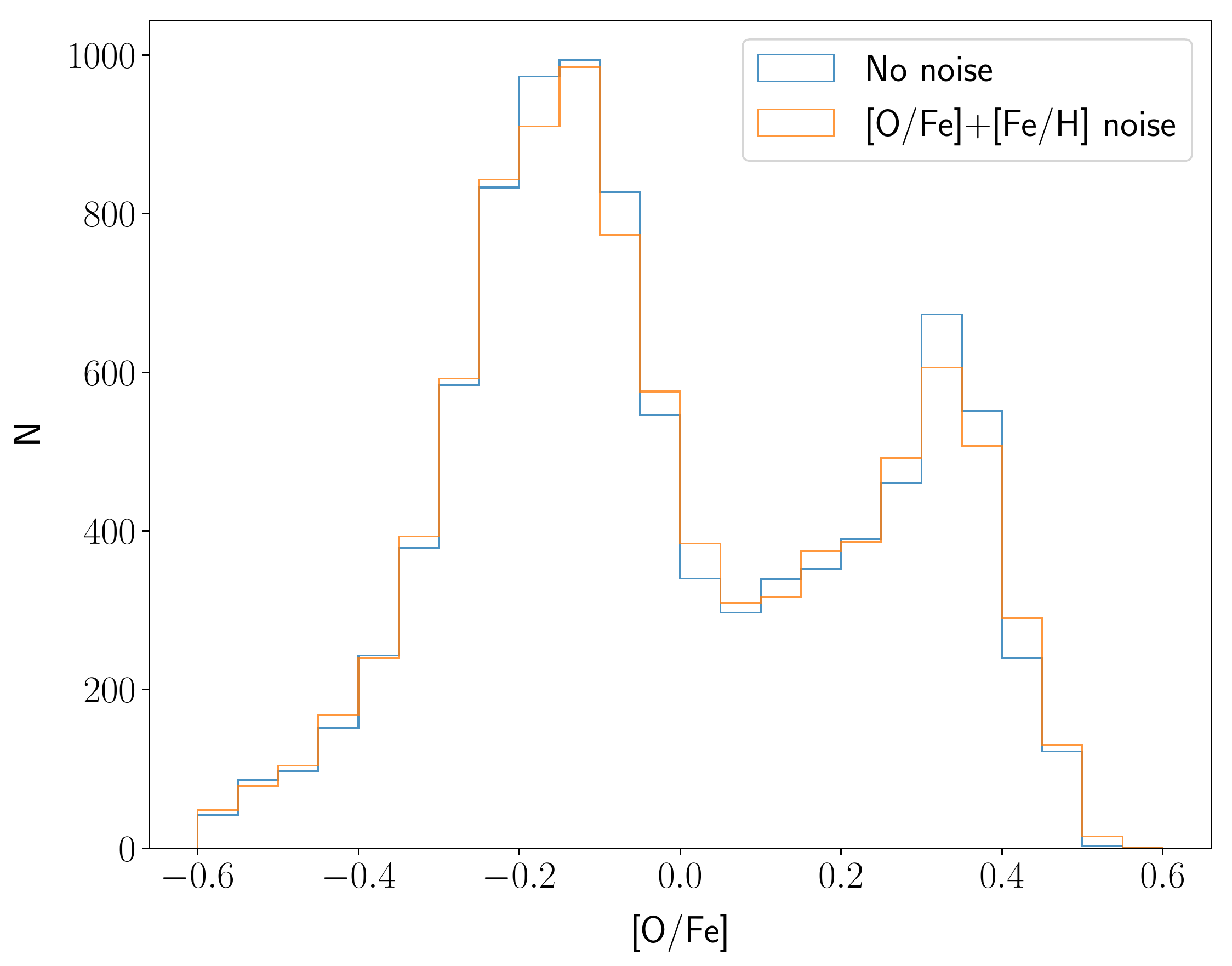}
        \caption{Histograms of \ofe\ for the sample at $-0.5 < \feh
        <-0.4$.  
	The cyan histogram shows the raw
        distribution from the model while the orange histogram shows
        the bimodality persists when noise typical of APOGEE is added
        to each star's chemistry.}
\label{f:double_peak}
\end{figure}

Figure \ref{f:radialcuts} shows the final ($10\Gyr$) distribution of
stars in the \ofe-\feh\ (chemical) plane at different radii and
heights.  As in the APOGEE survey \citep{Anders2014, Nidever2014,
Hayden2015} and other datasets \citep[e.g.,][]{Fuhrmann1998,
Bensby2005, Reddy2006, Adibekyan2012}, we find two distinct sequences,
a high-\al\ and a low-\al\ one.  The low-\al\ sequence dominates over
the high-\al\ one close to the mid-plane, while the high-\al\ sequence
is more prominent at larger heights.
The two sequences are bridged by stripes which are not observed in the
MW.  If we convolve the \ofe\ and \feh\ of stellar particles in the
Solar neighbourhood (defined throughout this paper as $7 < R/kpc < 9$)
with the APOGEE observational errors, $\sig{\feh} = 0.1$ dex and
$\sig{\ofe} = 0.03$ dex \citep{Nidever2014} the stripes visible in
Figure \ref{f:radialcuts} are masked, without erasing the chemical
bimodality.  Figure \ref{f:double_peak} directly shows the bimodality
in \ofe\ at the Solar neighbourhood for the cut through the chemical
space $-0.5 < \feh < -0.4$.  The distribution is bimodal, with a very
clearly defined minimum, even when the distribution is convolved with
the observational errors, as it must be.

We define two lines by eye to separate the chemical space into
low-\al\ and high-\al\ sequences:
\begin{equation}
\ofe <
\begin{cases}
 0.3 & \text{if \feh\ $< -0.8$} \\
 -0.3913 \feh - 0.0130 & \text{otherwise.}
\end{cases}
\label{e:sequences}
\end{equation}
Throughout this paper, we refer to stars above this locus as the
high-\al\ sequence and those below as the low-\al\ sequence.  These
lines are overplotted on Figure \ref{f:radialcuts}.  Across the entire
galaxy, the mass of the high-\al\ sequence is $1.7 \times10^{10}
\Msun$ out of a total stellar mass of $6.23 \times10^{10} \Msun$,
\ie\ the high-\al\ sequence contains $28\%$ of the stellar mass, which
decreases to below $20\%$ if only the mass outside $3\kpc$ is
considered.  In comparison, \citet{snaith+14} estimate that the
entire chemically-defined thick disc of the MW constitutes $47\%$ of the
stellar mass.

\begin{figure*}
\centerline{
\includegraphics[angle=0.,width=0.5\hsize]{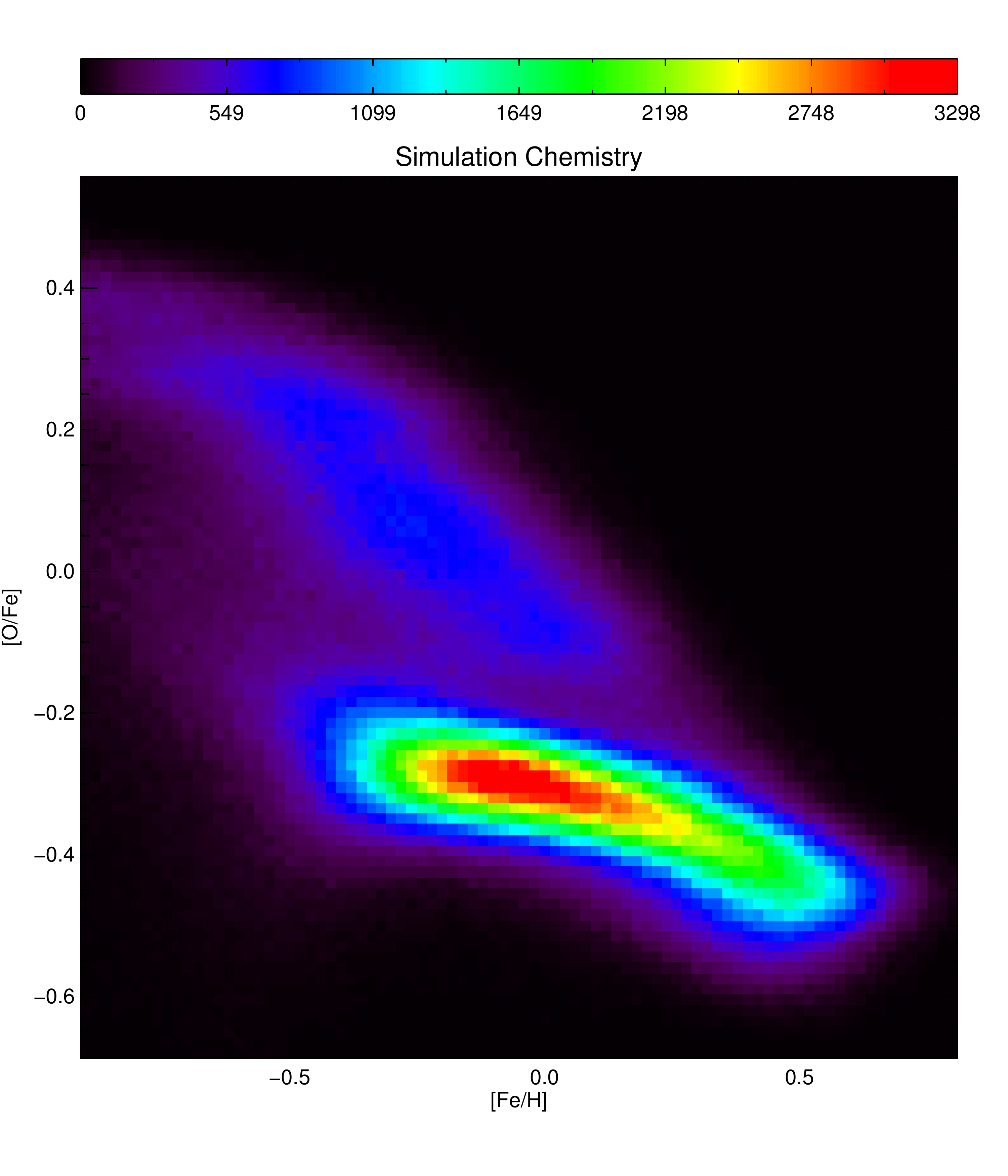}
\includegraphics[angle=0.,width=0.5\hsize]{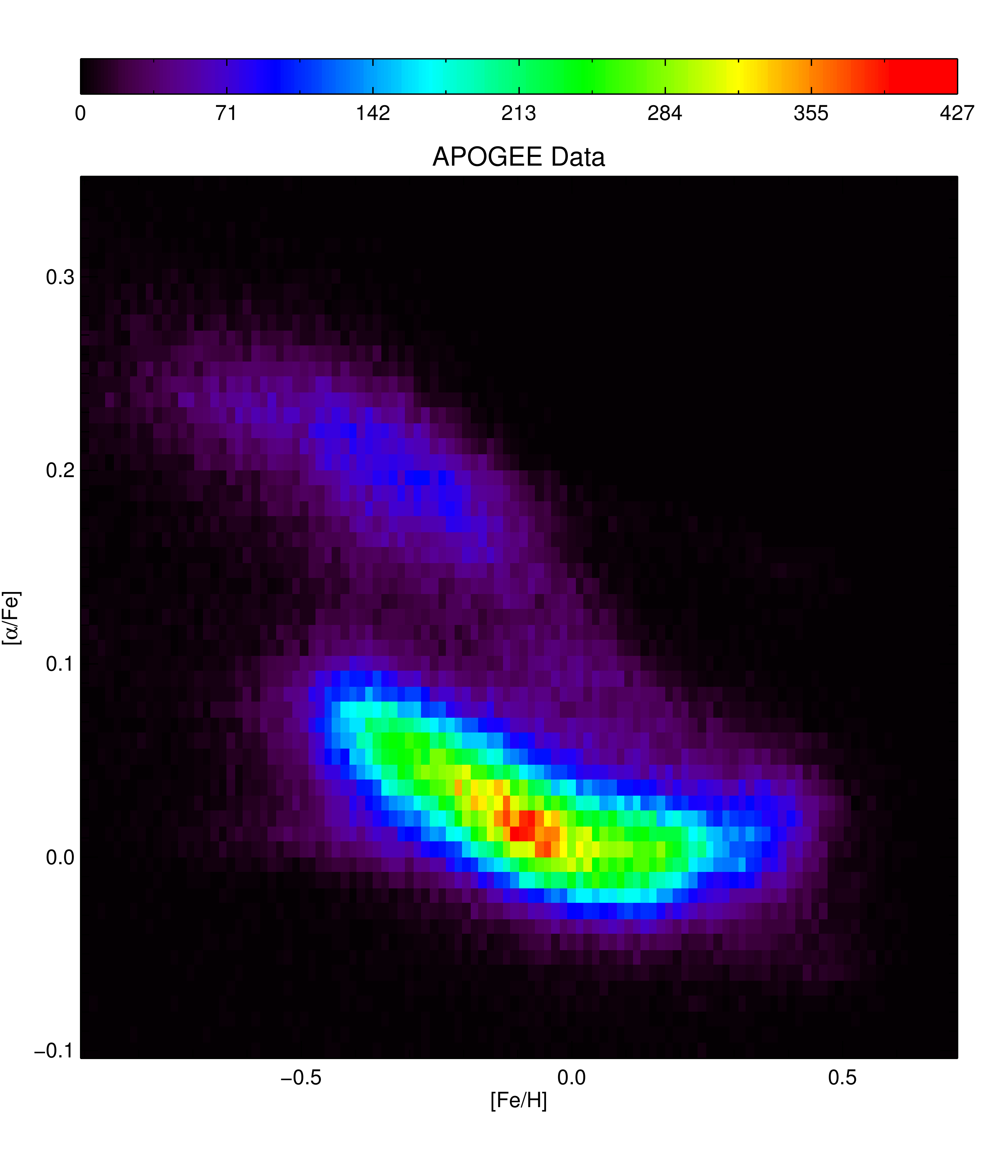}
}
\caption{Chemical comparison with APOGEE data.  On the left is the
  model chemistry in the \ofe-\feh\ space convolved with APOGEE-like
  uncertainties, while the right presents the chemistry of red giant
  stars in the APOGEE survey (DR14) in the \alfe-\feh\ space.}
\label{f:APOGEEComp}
\end{figure*}

Figure \ref{f:APOGEEComp} compares the chemical distribution in the MW
as observed by APOGEE throughout a large extent of the galaxy
\citep[DR14 red giants,][]{Holtzman2018} with that of the model 
convolved with the APOGEE observational errors.  The lower sequence
has the opposite curvature compared with the model but this may be
sensitive to the star formation history and the specific elements used
\citep[e.g.][]{Ramirez2007}, as is the scale of the vertical axis.
The \alfe\ defined by APOGEE includes oxygen but also other
\al-elements; when we consider just \ofe\ in the APOGEE data, although
the two sequences are noisier, the curvature of the low-\al\ track is
similar to the simulation's at the low \feh\ end, though still
opposite at high \feh.  Nonetheless, an unambiguous and sustained
separation into two tracks is readily apparent in the model, as in the
MW.  Additionally, the extent and curvature of the upper sequence is
similar to that in the APOGEE data.  The facts that the high-\al\
sequence covers nearly the same \feh\ range as the low-\al\ sequence,
and that the two sequences approach each other at the high \feh\ end,
are especially intriguing constraints on the formation of the
bimodality because it implies that it did not form in a single
discrete event such as infall of pristine gas
\citep[e.g.][]{Mackereth2018}.  The extension of the high-\al\
sequence to lower metallicities than the low-\al\ also matches the MW.
Still, the model and the MW do not match in detail; one of the more
significant differences is the larger separation between the high-\al\
and the low-\al\ sequences.  This may be a problem of either the
yields used (also supported by the relative vertical offset of the
model and the MW) or the star formation rates that produce the
high-\al\ sequence.

\subsection{Present day properties of the high-\al\ sequence}

\begin{figure*}[!h]
	\hskip -.3in \epsscale{1} \plotone{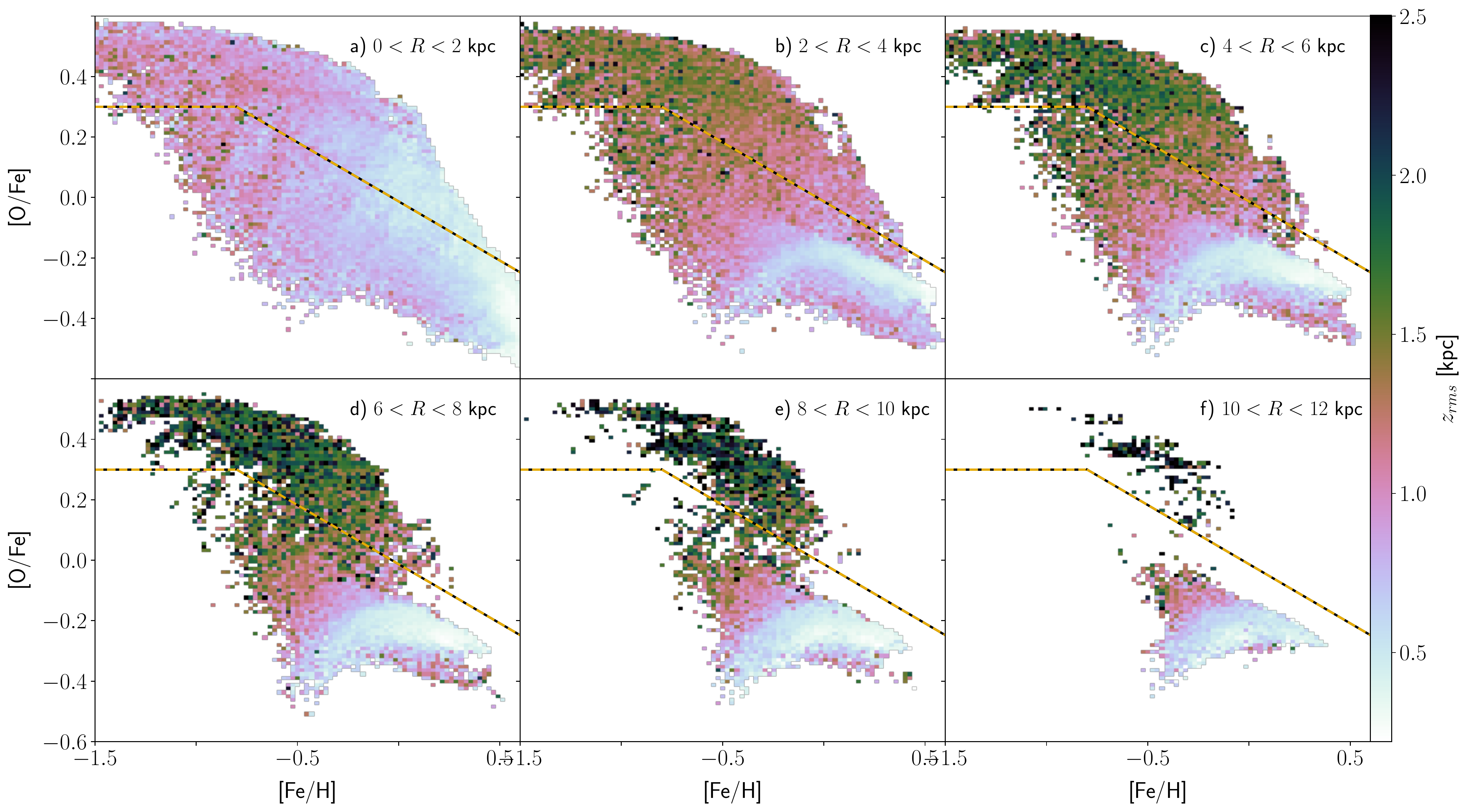}
	\caption{Root-mean-square height, $z_{\textrm{rms}}$, of stars
	in chemical space for various radial cuts, as labelled in each
	panel. We suppress bins with less than 10 star particles.  The
	dashed yellow-black line, defined by Eqn. \ref{e:sequences},
	separates the two sequences.}  \label{f:zrms}
\end{figure*}

\begin{figure*}[!h]
	\hskip -.3in \epsscale{1}
	\plotone{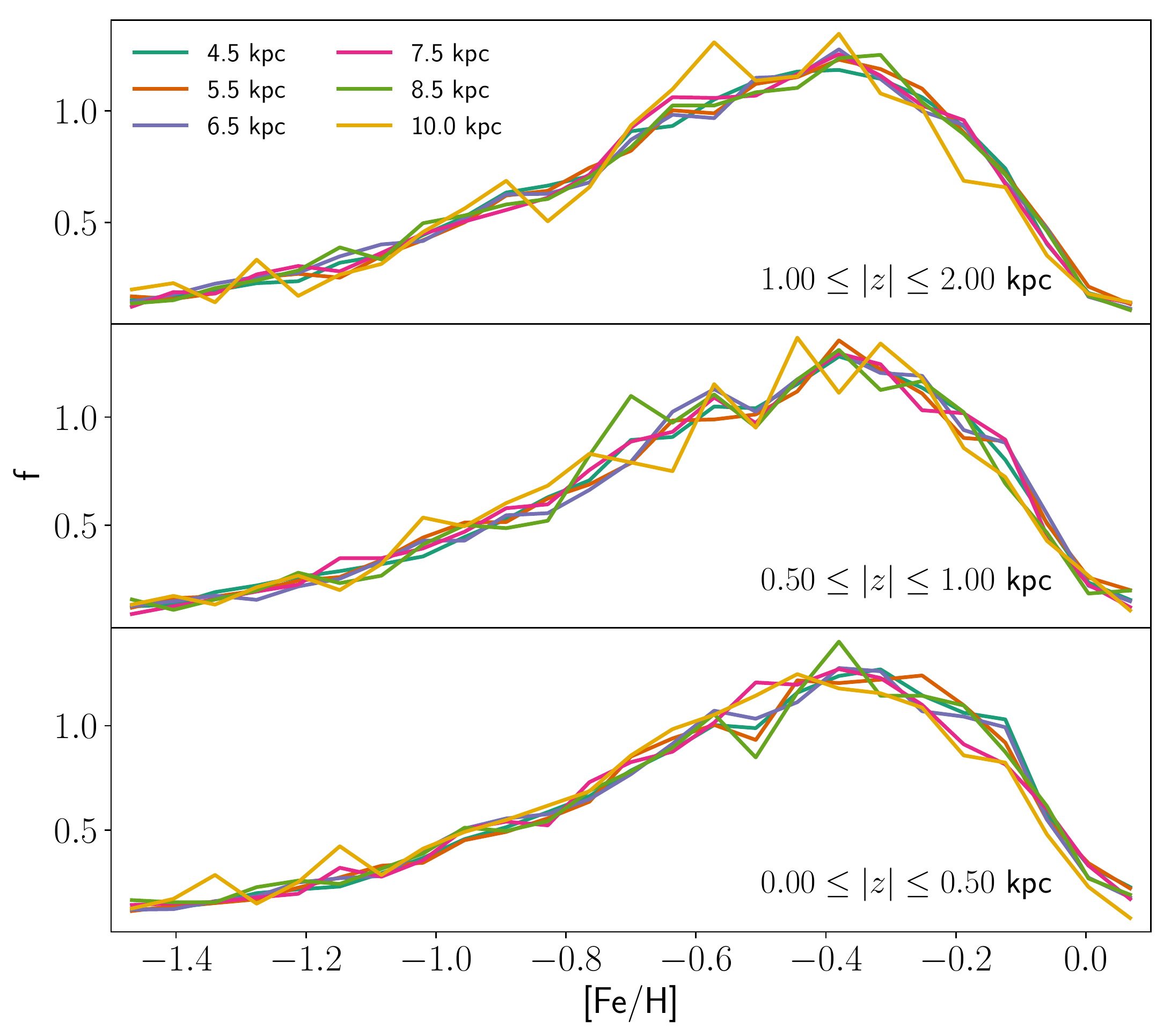}
	\caption{MDFs of the \al-enhanced stars (here defined as
	$\ofe> 0.2$) at different radii (indicated by different
	colours) and different heights above the mid-plane (as
	indicated in each panel).  The MDFs of this population are
	homogenous across the galaxy. } \label{f:mdf-radial}
\end{figure*}

\begin{figure}[!h]
	\hskip -.3in \epsscale{1}
	\plotone{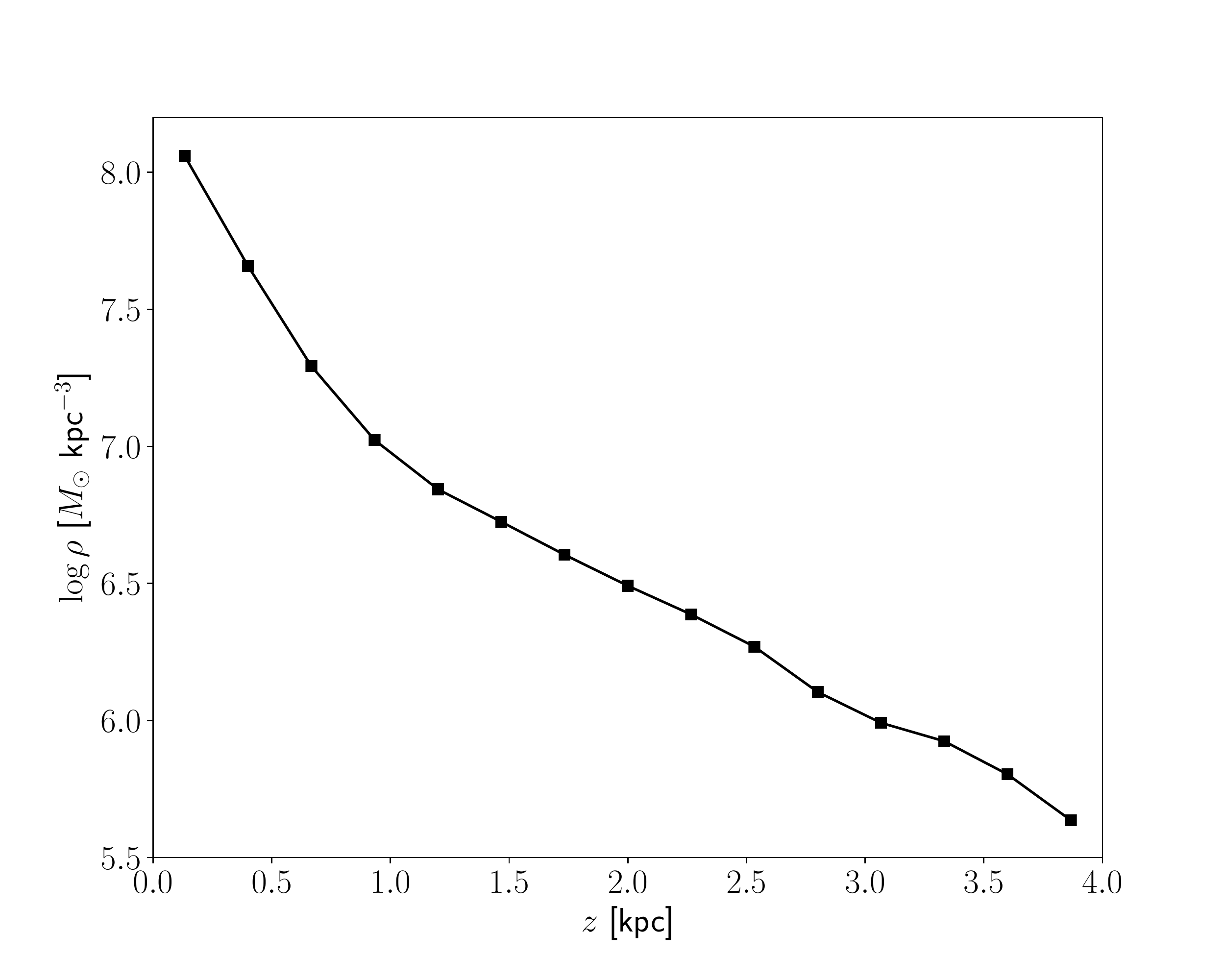} 
	\caption{The model's vertical density distribution in the
	Solar neighbourhood.  } 
\label{f:vertprof}
\end{figure}

The spatial distribution of the high-\al\ stars is also evident in
Figure \ref{f:radialcuts}. High-\al\ stars are more prominent at large
heights at the present time.  The high-\al\ distribution is also more
centrally concentrated than the low-\al\ ones; in the MW
\citet{Bovy2012b} found shorter scale-lengths for low-\al\ stars while
\citet{Hayden2015} found little evidence of the high-\al\ sequence
outside the Solar neighbourhood.  Figure \ref{f:zrms} shows the
present day root mean square vertical height of stars across the
chemical space.  The high-\al\ stars are distributed in a thicker disc
than the low-\al\ stars outside the bulge region
\citep[e.g.][]{Bovy2012b}.  The variation of height is continuous
across the chemical space, as found in the MW by \citet{Bovy2012a}.
This important constraint suggests that the high-\al\ stars are not
the product of a single violent event which had suddenly heated
the disc vertically.  In spite of this continuous height variation
the vertical stellar density profile is very clearly
double-exponential, as can be seen in Fig. \ref{f:vertprof} for the
Solar neighbourhood (other regions in the disc look qualitatively
similar).  Fitting two exponentials to the Solar neighbourhood
vertical profile of the model gives scale-heights $h_1 = 240 \pm 8\pc$
and $h_2 = 1083 \pm 52\pc$; in comparison \citet{Juric2008} measured
$h_1 = 300\pm 50\pc$ and $h_2 = 900 \pm 180\pc$ in the Solar
neighbourhood. The geometric thin and thick discs have a transition
at about $1\kpc$, which is comparable to the transition in the MW
\citep[e.g.][]{Gilmore1983}.

\begin{figure*}[!h]
	\hskip -.3in \epsscale{1} \plotone{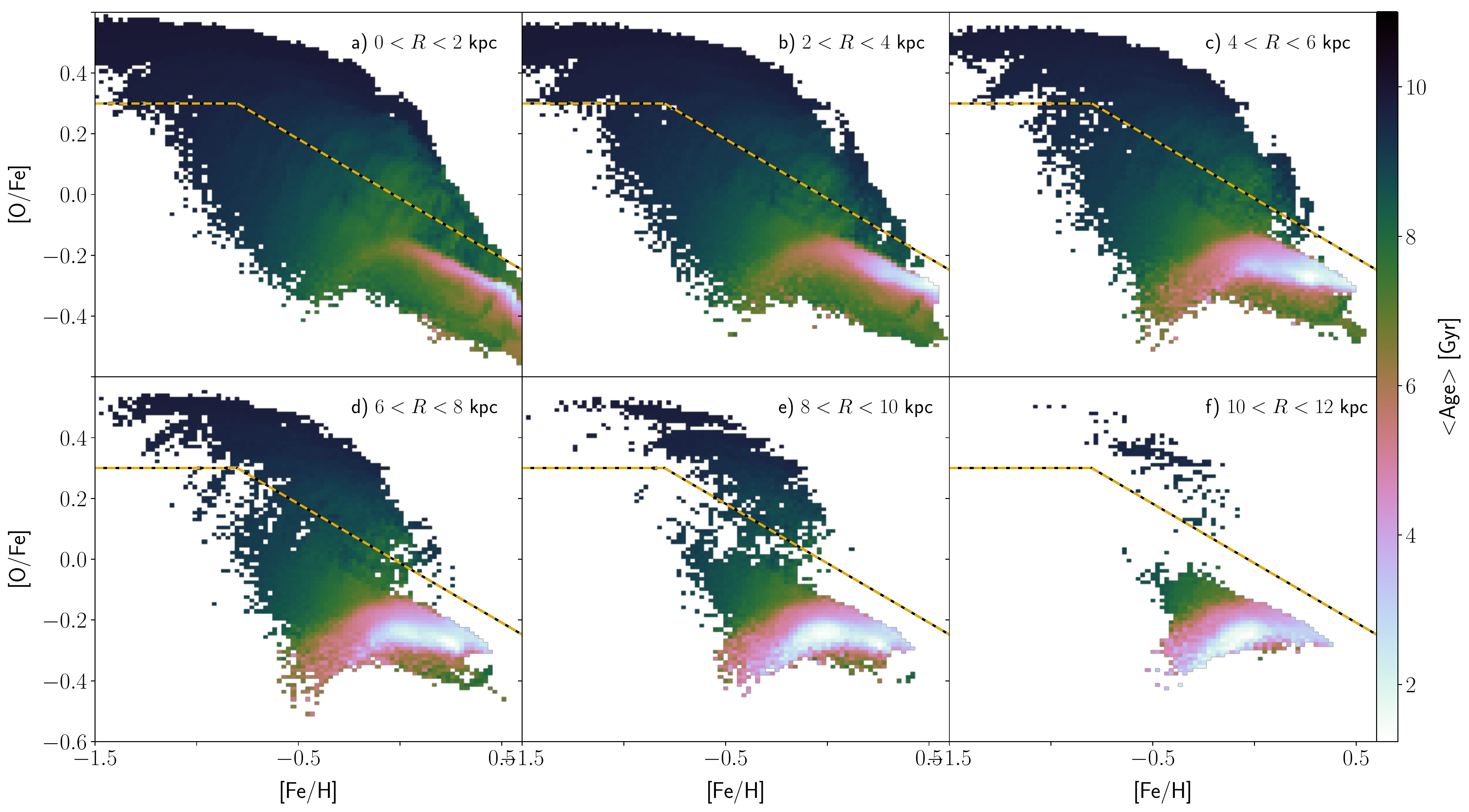}
	\caption{Mean age of stars in the chemical space for various
	radial cuts, as labelled in each panel. We suppress bins with
	less than 10 star particles.  The dashed yellow-black line,
	defined by Eqn. \ref{e:sequences}, separates the two
	sequences.}  \label{f:age}
\end{figure*}

\begin{figure*}[!h]
	\hskip -.3in
	\epsscale{1}
	\plotone{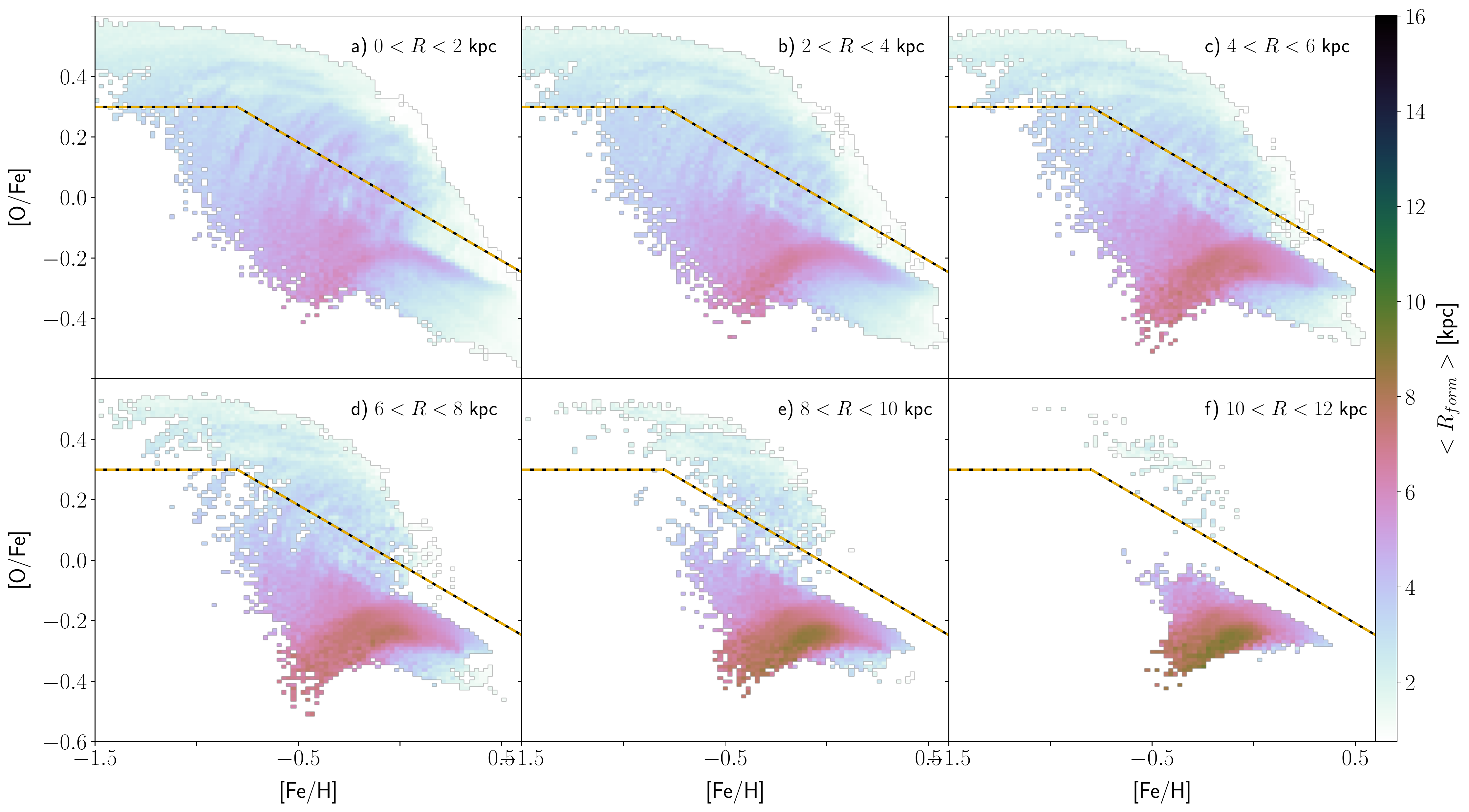}
	\caption{Mean radius of formation, $R_{\textrm{form}}$, of
          stars in the chemical space for various radial cuts, as
          labelled in each panel. We suppress bins with less than 10
          star particles.  The dashed yellow-black line, defined by
          Eqn. \ref{e:sequences}, separates the two sequences.}
	\label{f:rform}
\end{figure*}

\citet{Nidever2014} and \citet{Hayden2015} showed that the metallicity
distribution function (MDF) of \al-enhanced stars is essentially
independent of the location within the MW's disc.  Figure
\ref{f:mdf-radial} presents the MDFs of stars with $\ofe > 0.2$ 
\citep[note this \al-enhanced population is {\it not} identical to 
the high-\al\ sequence, but is chosen in the same spirit as
in][]{Nidever2014}, in the simulation for various radial and height
cuts. The overall shape and width of these MDFs are similar in all
regions considered.
\citet{Nidever2014} and \citet{Hayden2015} suggested that the similarity of 
the MDFs of \al-enhanced stars indicates these stars formed from
well-mixed gas.  \citet{Loebman2016} used simulations to demonstrate
that the similarity of the MDF of \al-enhanced stars is due to them
having formed in a relatively small region in the inner Galaxy over a
short time interval ($\la 2\Gyr$) and were subsequently migrated
throughout the disc.
Figure \ref{f:age} presents the average age of stars in the chemical
space.  The high-\al\ stars are old, having mean age $> 8 \Gyr$,
whilst the low-\al\ stars, particularly those at high \feh, are
predominantly young to intermediate age.  Figure \ref{f:rform}
presents the mean formation radius, $R_{\mathrm form}$, across the
chemical space.  The high-\al\ stars largely formed within the
inner $2 \kpc$, while the low-\al\ stars formed over a larger radial
region.  In agreement with \citet{Loebman2016}, therefore, we find
that the high-\al\ stars, and especially the \al-enhanced stars,
formed over a short time interval in a relatively restricted volume.

\begin{figure}
\centerline{
\includegraphics[angle=0.,width=\hsize]{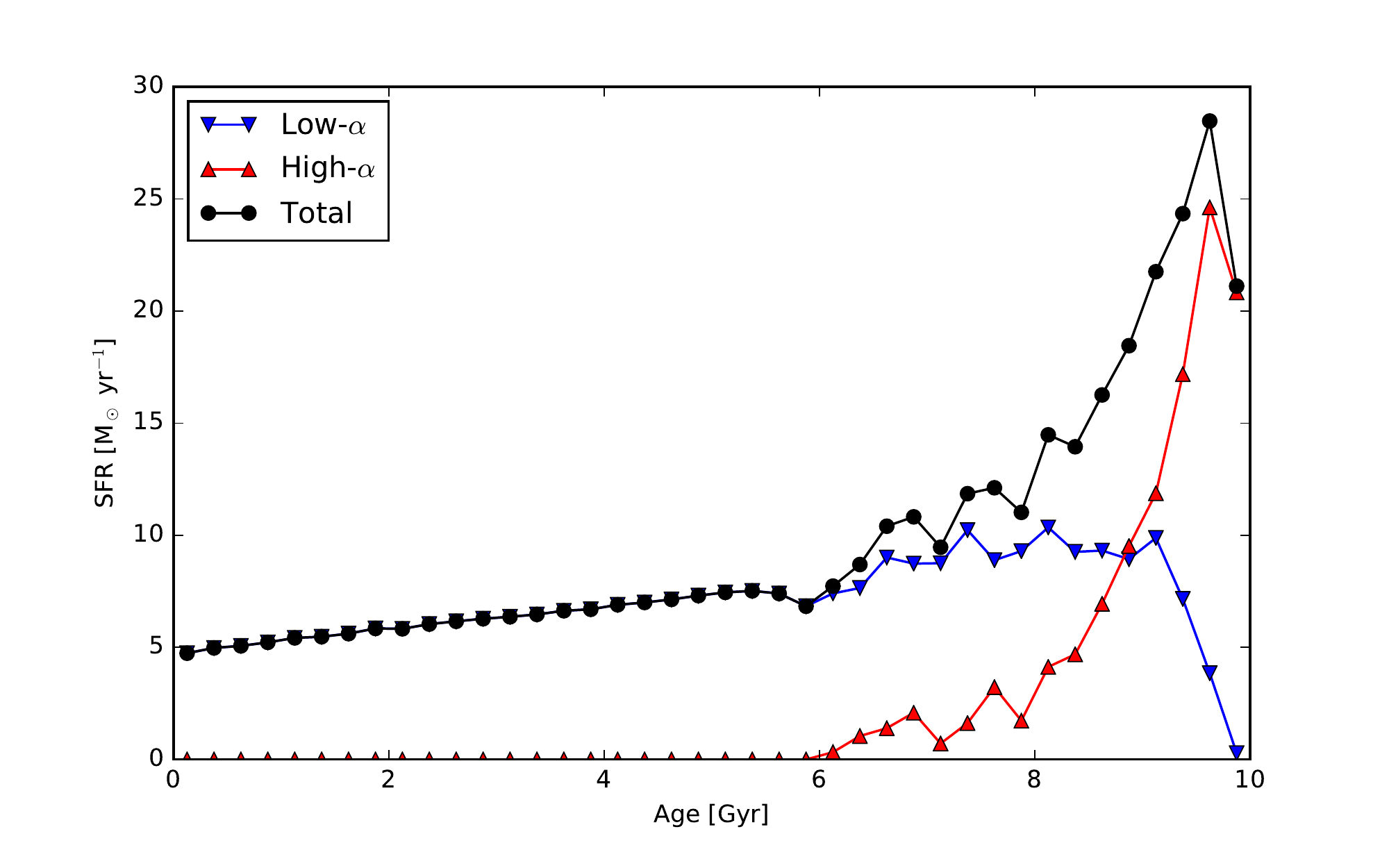}
}
\caption{The star formation history of the two sequences
  across the entire model.  The (red) triangles show the high-\al\
  sequence while the (blue) downwards-pointing triangles show the
  low-\al\ sequence.  The (black) circles show the total star
  formation history.}
\label{f:sfh}
\end{figure}

Figure \ref{f:sfh} shows the star formation history (SFH) of the
high-\al\ and low-\al\ stars.  The two populations overlap at old
ages, with the SFH of the high-\al\ stars peaking very early,
persisting to $\sim 4\Gyr$, but largely over by $2\Gyr$.  The SFH of
the low-\al\ stars is relatively flat after $1\Gyr$, becoming the main
mode of star formation thereafter.  The early formation of the
majority of the high-\al\ stars ensures that they are more centrally
concentrated then the subsequent low-\al\ disc stars, in good
agreement with what is found in the MW \citep{Bovy2012b, Bovy2016}.

\begin{figure*}
\centerline{
\includegraphics[angle=0.,width=0.5\hsize]{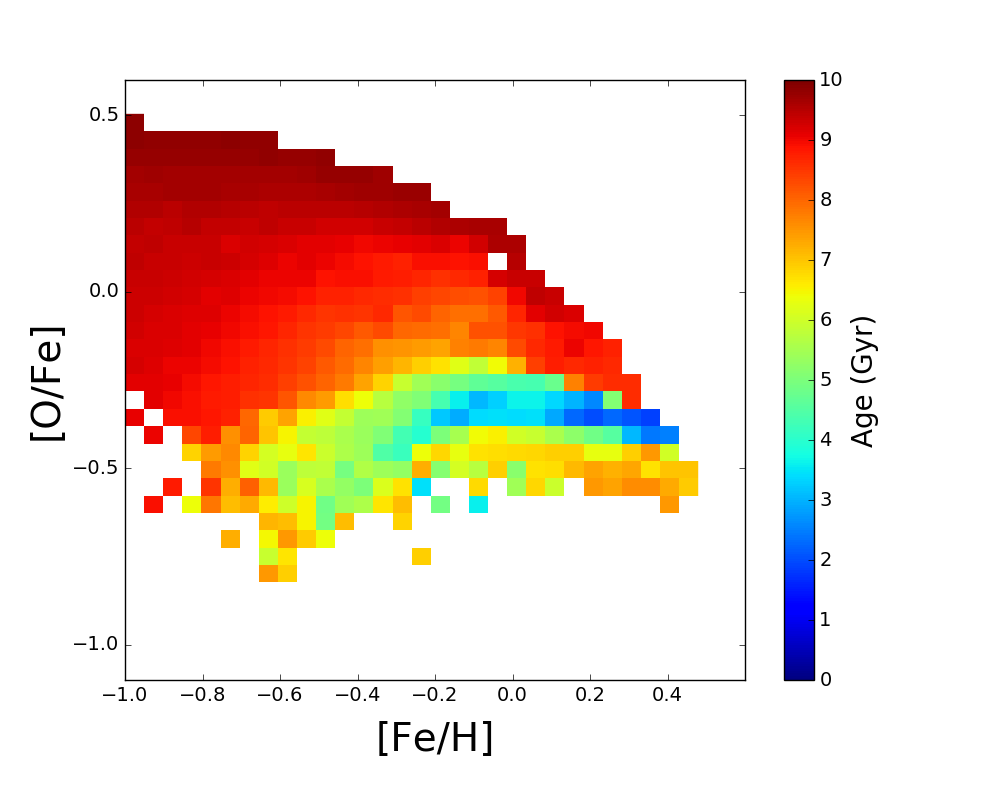}
\includegraphics[angle=0.,width=0.5\hsize]{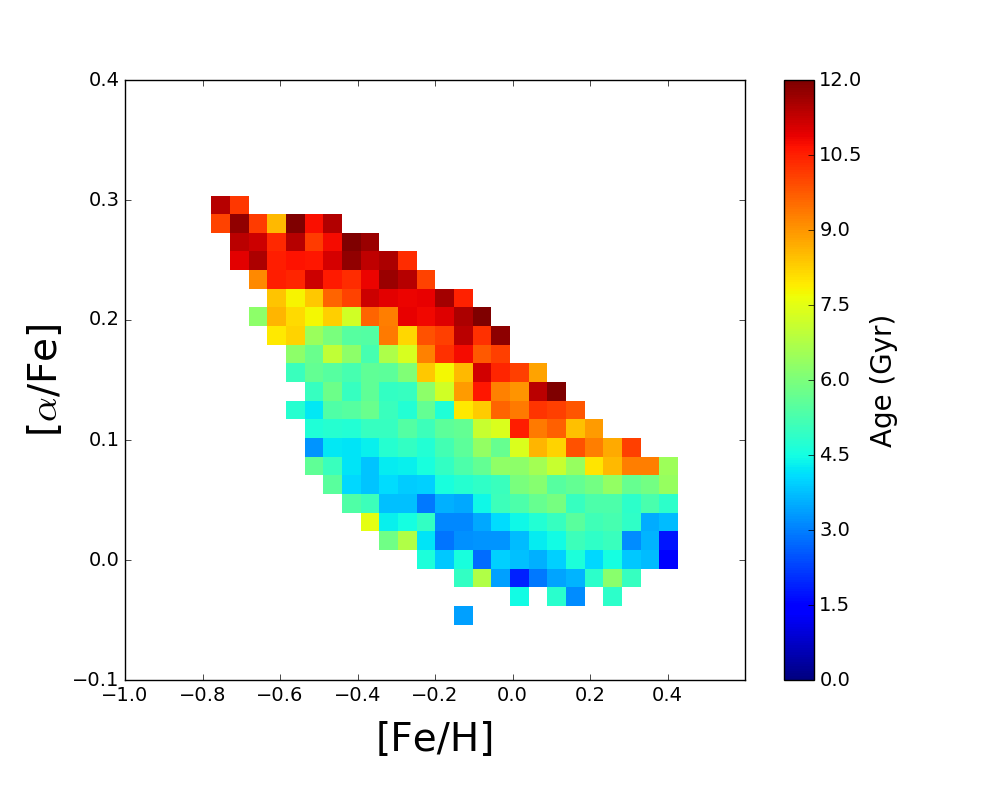}
}
\caption{Age comparison with APOGEE data.  On the left is the model's
  mean stellar age in \ofe-\feh\ space, while the right presents the
  mean age of stars in APOGEE in the \alfe-\feh\ space.}
\label{f:APOGEEAges}
\end{figure*}

Figure \ref{f:APOGEEAges} compares the stellar ages in the MW from the
APOGEE survey and the simulation.  The left panel, showing the ages in
the simulation, uses star particles in the simulation with $|z| < 1
\kpc$, and across Galactic radii similar to the red clump 
($3 < x < 13\kpc$ and $|y| > 5\kpc$).  The right panel uses $\sim
20,000$ red clump stars in APOGEE with ages from \citet{Ness2016}.
While ages of the high-\al\ stars are similar between the model and
the MW, the two are quite different in detail.  The biggest difference
is at low metallicity, where the model has young stars only at low
\al.  The model's chemical enrichment history therefore is different
from that of the MW.  This makes the similarities in the bimodality
even more remarkable.

\subsection{Two modes of star formation}

We now investigate in greater detail the origin of the high-\al\
sequence.  High-\al\ stars are generally thought to form at high star
formation rate (SFR); we compute the local SFR density,
$\Sigma_\mathrm{SFR}$, at birth of each star particle from the
distribution of stars forming in each $10 \Myr$ interval.  We estimate
the density of these new stars by recovering their positions at the
beginning of the time interval using the mean disc rotation velocity
at their radius.  We then calculate the density of the particles in
the disc plane by binning the particles in Cartesian coordinates with
a bin size of 100 pc $\times$ 100 pc.  The grid is then smoothed with
a Gaussian of FWHM$=200 \pc$ and the values are converted to physical
units of $\Msun \yr^{-1} \kpc^{-2}$.  Finally we assign local
$\Sigma_\mathrm{SFR}$ values to each star particle on the basis of the
bin within which it falls.

\begin{figure*}
\centerline{
\includegraphics[angle=0.,width=0.9\hsize]{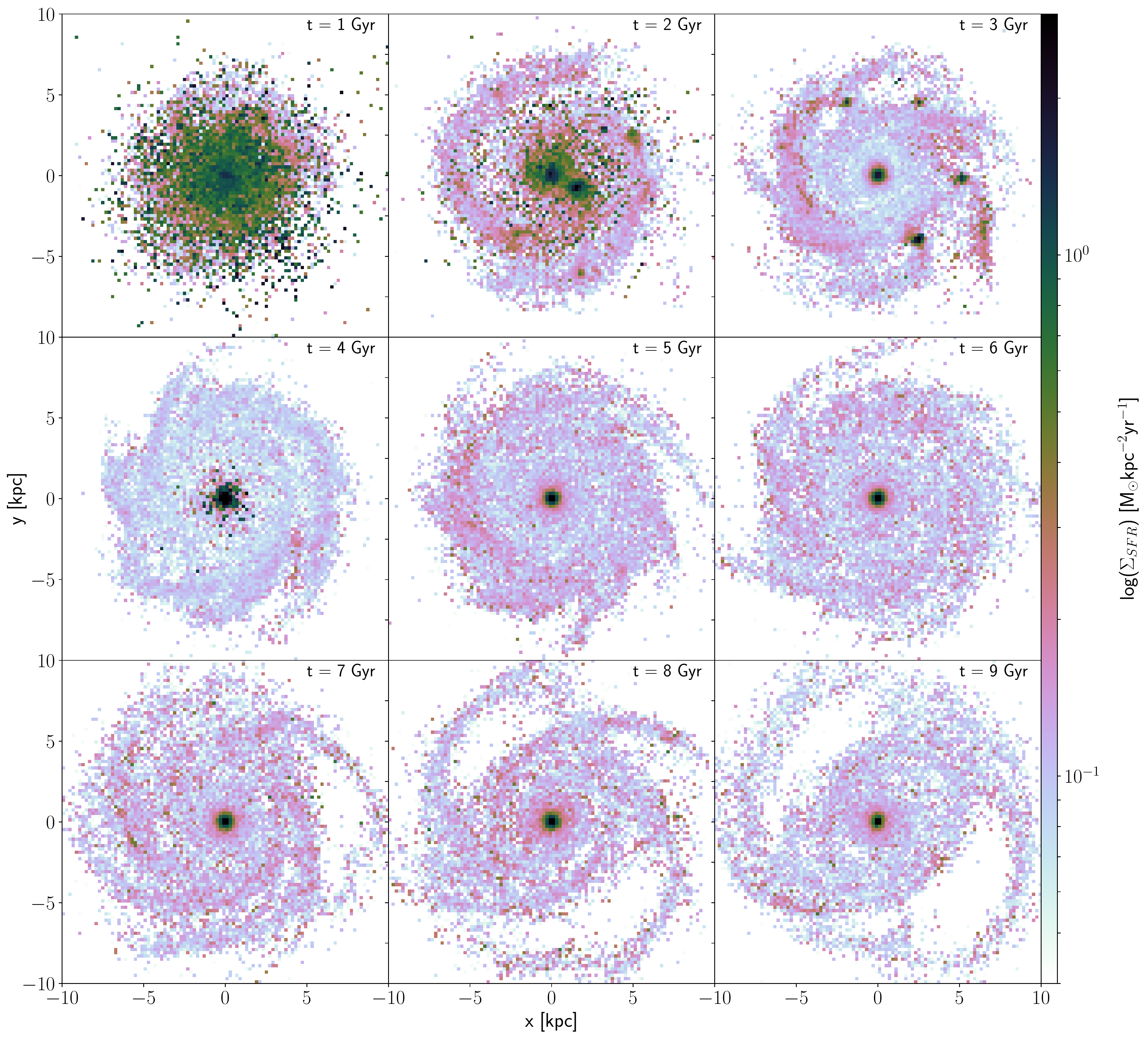}
}
\caption{Maps of the star formation rate computed as described in the
  text.  The panels are spaced by $1 \Gyr$ from $t=1\Gyr$ (top left)
  to $9\Gyr$ (bottom right). While low $\Sigma_{SFR}$ ``smooth'' star
  formation occurs throughout the galaxy at all times, there is high
  $\Sigma_{SFR}$ clumpy star formation only during the first few Gyr.
  Each panel shows the distribution of stars younger than $200\Myr$;
  during this time some high-\al\ stars will have diffused out of
  their birth clumps. This is particularly evident at $t=1\Gyr$.}
\label{f:SFRmaps}
\end{figure*}

Figure \ref{f:SFRmaps} presents snapshots of the local star formation
rate computed this way over the course of the simulation.  At early
times a considerable amount of star formation occurs in clumps, which
have a high SFR density. As in the two-infall model of
\citet{Chiappini1997} this mode of star formation becomes negligble
after a few Gyr, aside from at the bulge.  At the same time as the
clumpy star formation is occurring, stars are forming in the disc in a
more distributed (``smooth'') configuration with much lower
$\Sigma_\mathrm{SFR}$.  This is the main mode of star formation after
$3\Gyr$.

\begin{figure*}[!h]
	\hskip -.3in \epsscale{1}
	\plotone{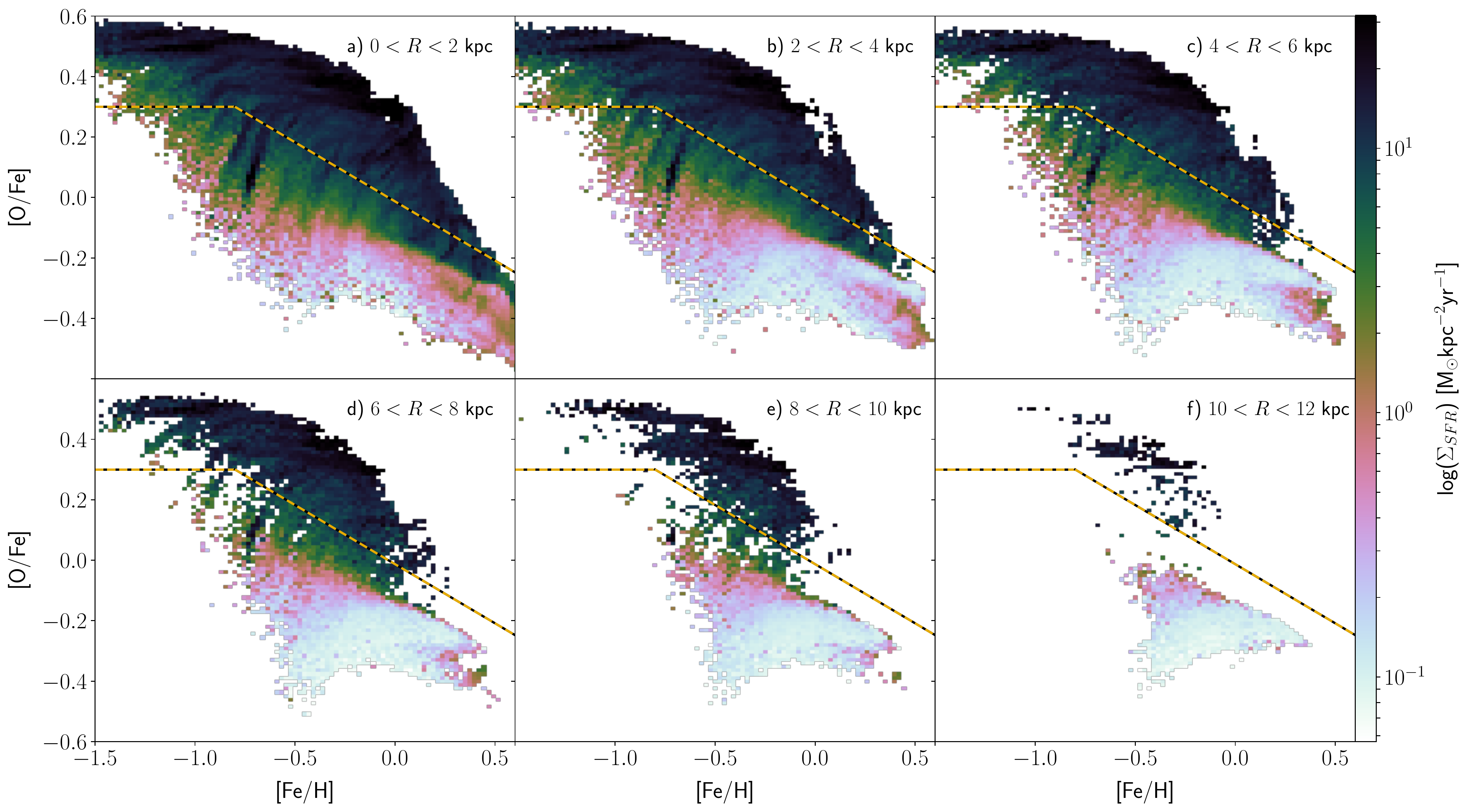} \caption{The mean
	$\Sigma_\mathrm{SFR}$ of star particles in the chemical
	abundance plane in various spatial bins.  The high-\al\ stars
	have high $\Sigma_\mathrm{SFR}$ while low-\al\ stars have low
	$\Sigma_\mathrm{SFR}$.  The dashed yellow-black line, defined
	by Eqn. \ref{e:sequences}, separates the two sequences.  We
	suppress bins with less than 10 star particles.}
	\label{f:localsfr}
\end{figure*}

\begin{figure}[!h]
	\hskip -.3in \epsscale{1.0}
	\plotone{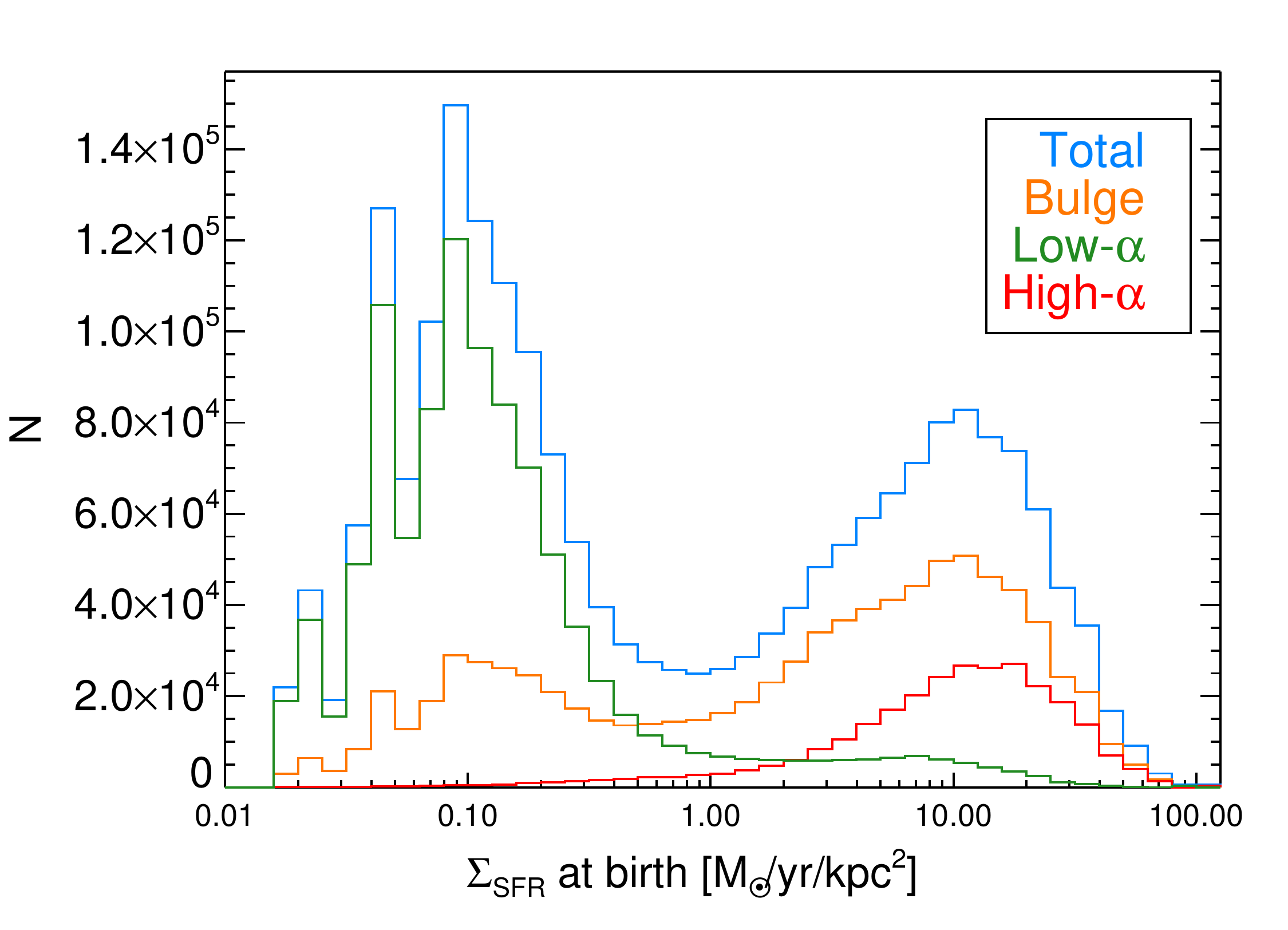} \caption{Histogram of
	$\Sigma_\mathrm{SFR}$ for all star particles.  This shows a
	bimodality in $\Sigma_\mathrm{SFR}$ with a minimum at
	intermediate star formation rates.  The low-\al\ and high-\al\
	histograms are obtained for all stars outside $R > 2 \kpc$,
	whereas the bulge distribution is for stars at $R < 2 \kpc$.
	The bimodality is not produced by stars at the bulge.}
\label{f:localsfrhist}
\end{figure}

Figure \ref{f:localsfr} shows $\Sigma_\mathrm{SFR}$ of star particles
at the time of formation across the chemical plane.  The high-\al\
sequence has a significantly higher $\Sigma_\mathrm{SFR}$, which
results in being \al-rich, while the low-\al\ stars have much lower
local $\Sigma_\mathrm{SFR}$ and chemically evolve more slowly.
Unlike Figure \ref{f:zrms}, Figure \ref{f:localsfr} has an abrupt
transition between the high-\al\ and the low-\al\ sequences.  Figure
\ref{f:localsfrhist} presents histograms of $\Sigma_\mathrm{SFR}$ at
formation, showing that there is a bimodality in
$\Sigma_\mathrm{SFR}$.  The two modes of SF differ by a factor of
roughly 100 in $\Sigma_\mathrm{SFR}$.  The high-\al\ sequence forms
via the high-SFR density, clumpy star formation mode, while the
low-\al\ sequence forms via the low-SFR density, smooth star formation
mode that is dominant at later times.  As already shown in Figure
\ref{f:sfh}, the low-\al\ mode of star formation is already present at
early times at the same time as the high star formation rate,
high-\al\ mode.  This simultaneous formation of both \al-sequences,
via the dual-modes of star formation, appears to be in agreement with
the recent observational results of
\citet{SilvaAguirre2017} and \citet{Hayden2017b}.

\begin{figure*}
\includegraphics[angle=0.,width=\hsize]{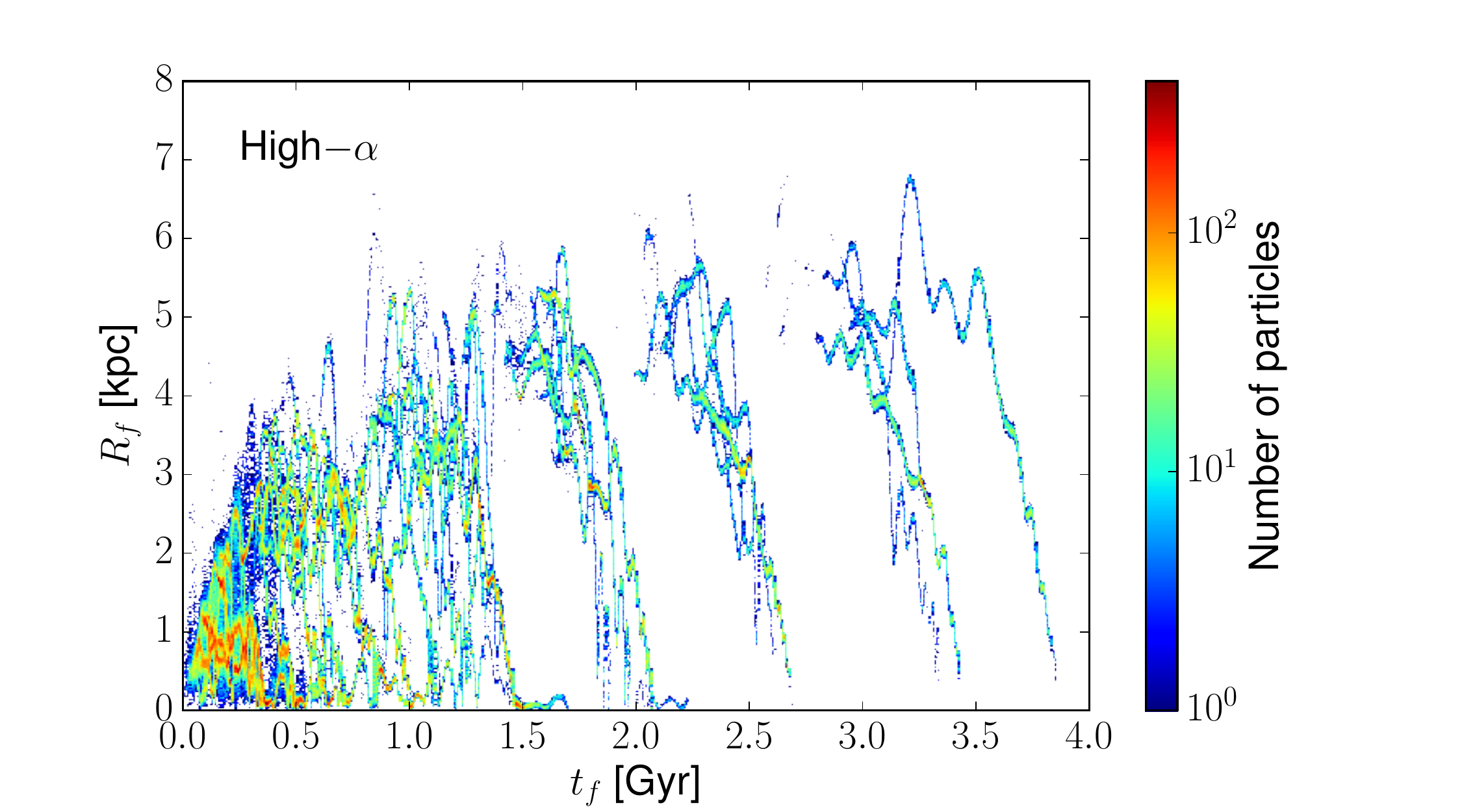}\\
\includegraphics[angle=0.,width=\hsize]{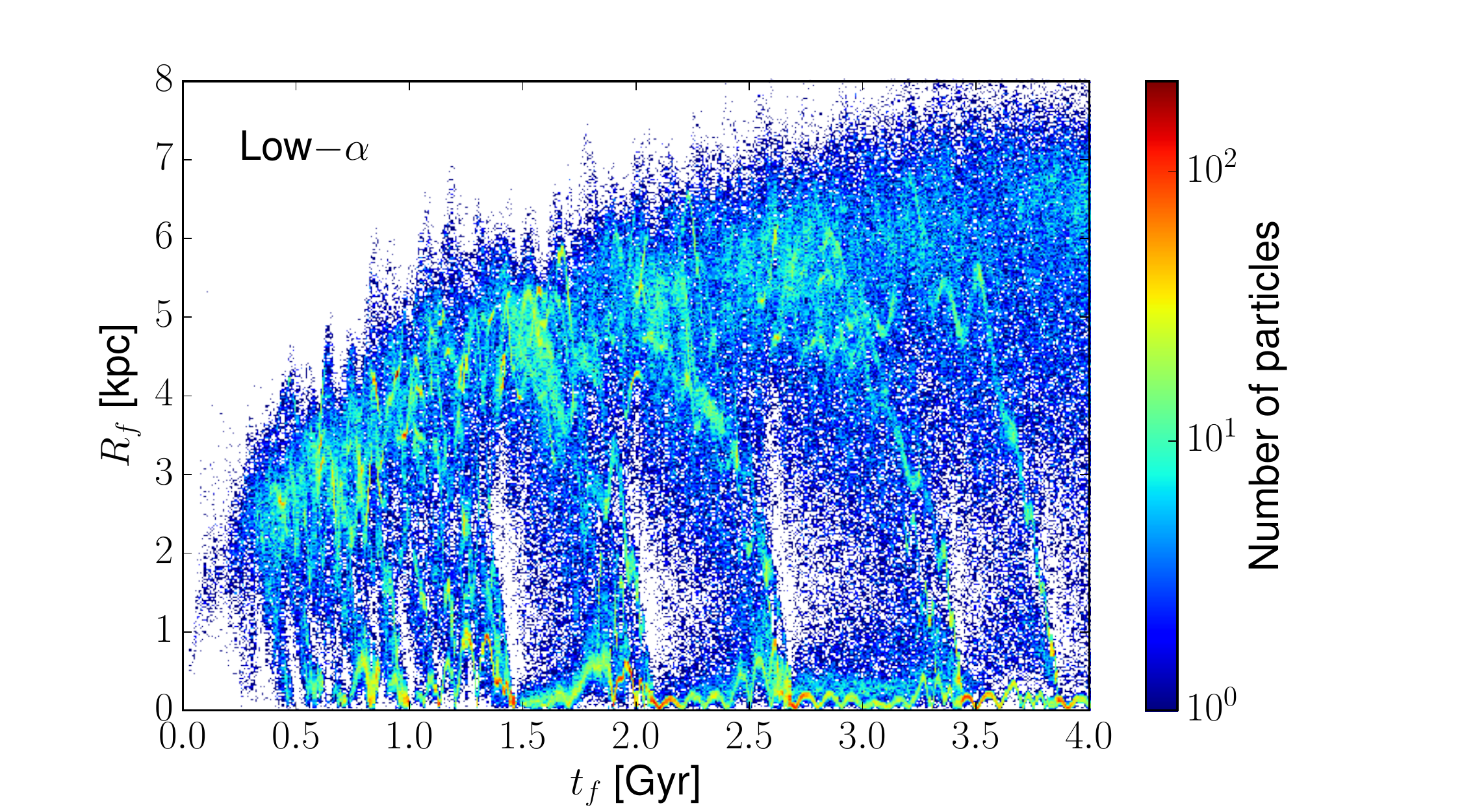}
\caption{The evolution of the radius at which stars form for the
  high-\al\ (top) and low-\al\ (bottom) sequences over the first
  $4\Gyr$.  }
\label{f:sflocation}
\end{figure*}

Figure \ref{f:sflocation} shows the evolution of the radius at which
stars are forming in the high-\al\ and low-\al\ sequences.  The
high-\al\ sequence is seen to derive from a decreasing number of
discrete star forming clumps that sink towards the centre.  The star
formation in the low-\al\ sequence is more broadly distributed,
although the effect of spirals is evident as gaps in the radial
distribution at $R_f < 3\kpc$ at certain times.  These spirals are
associated with the clumps, by being partly excited by them
\citep{DOnghia+13} and also partly needing the density enhancement 
provided by spirals to assemble the gas needed to form clumps.
Low-\al\ star formation is also associated with the clumps throughout
a large fraction of their trajectories, but the bulk of low-\al\ stars
form outside clumps, particularly after the first \Gyr.  A similar
plot to Figure \ref{f:sflocation} but separated into a low and high
SFR density at $\Sigma_\mathrm{SFR} = 30\Msun\kpc^{-2}\yr^{-1}$ is
virtually identical apart from having more star formation in the
bulge, persisting for the full $10\Gyr$.  About $43\%$ of stars form
in the high SFR density component, which reduces to $31\%$ if the
inner $0.5\kpc$ is excluded. In comparison we earlier found that the
high-\al\ sequence comprised $28\%$ of all stars. Comparing these two
numbers we conclude that the majority of high-\al\ stars form at high
$\Sigma_\mathrm{SFR}$ within clumps.

\begin{figure}
\includegraphics[angle=0.,width=\hsize]{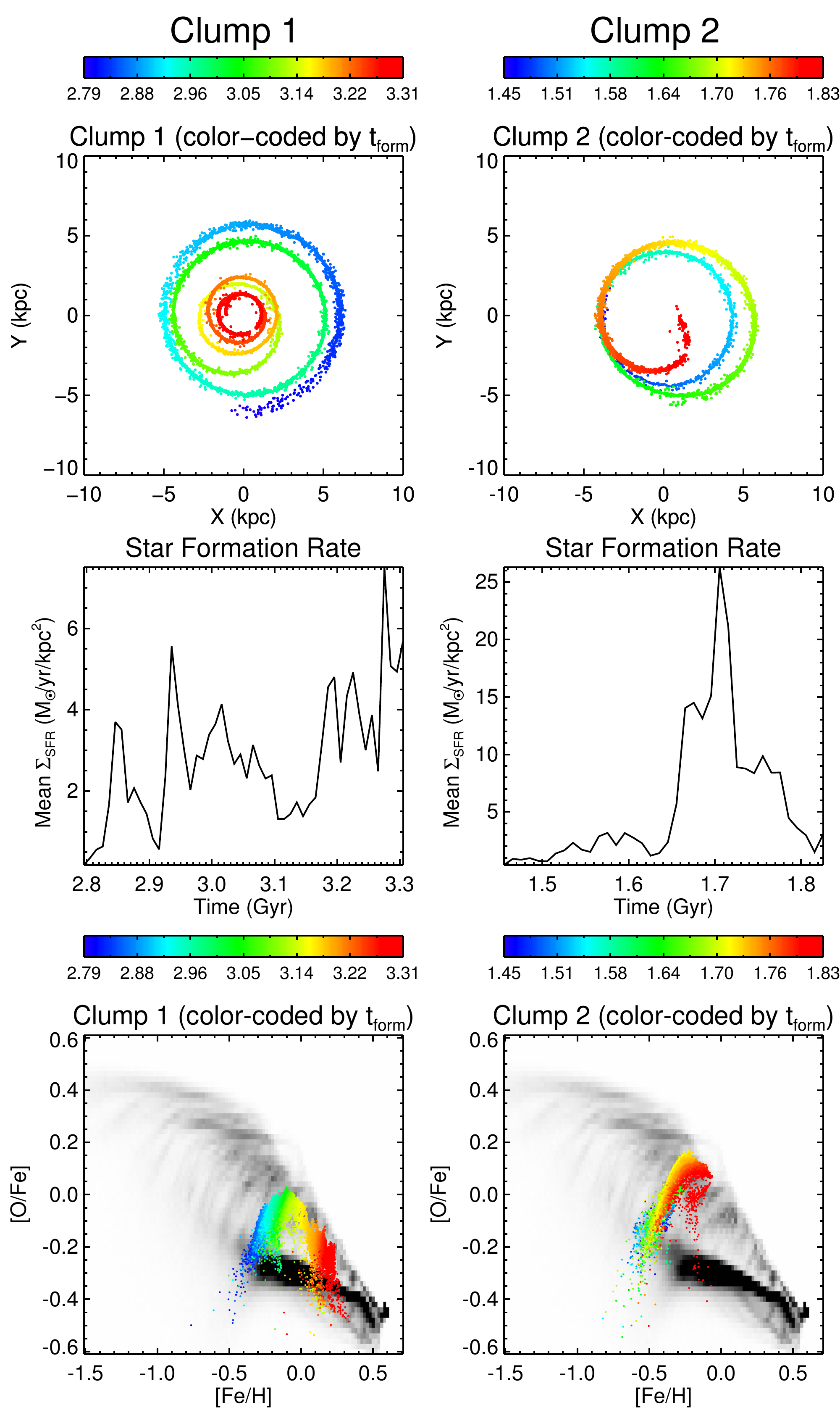}
\caption{The evolution of two clumps in configuration space (top),
star formation rate density (middle), and chemical space (bottom).
Clump 1 (left panels) forms around $2.8 \Gyr$ and over the next
$0.5\Gyr$ sinks to the centre.  During this time it quickly
transitions from the low-\al\ to the high-\al\ sequence.  Clump 2
(right panels) behaves similarly but towards the end of its life it
merges with another clump and quickly falls to the centre.  Colour
codes for the time, as indicated by the wedges.  In the bottom row,
the grey scale indicates the distribution of all stars at the end of
the simulation.}
\label{f:2clumps}
\end{figure}

Figure \ref{f:2clumps} presents the evolution of two typical clumps
chosen by eye.  Clump 1 forms at $\sim 6\kpc$; initially on the
low-\al\ sequence, it quickly shoots up to the high-\al\ sequence as
$\Sigma_\mathrm{SFR}$ rises, becoming more \feh-rich along this
sequence as supernovae of type Ia commence.  Its orbit decays and by
the time it reaches $1\kpc$ it starts falling back on to the low-\al\
sequence.  At this point the clump is lost in the bulge.  Clump 2 also
ascends from the low-\al\ to the high-\al\ sequence.  Its orbital
radius is rather more constant, and it evolves in \feh\ less over its
shorter lifetime ($300\Myr$ versus $500\Myr$ in clump 1).  Towards the
end of its life it merges with another clump and quickly drops to the
centre of the galaxy, dropping from the high-\al\ to the low-\al\
sequence.  These two clumps illustrate an important point: contrary to
the usual interpretation, the high-\al\ sequence is reached {\it from}
the low-\al\ sequence, rather than forming before it.  These examples
also demonstrate that the stripes connecting the low-\al\ to the
high-\al\ sequence are due to the evolution of individual clumps in
chemical space.

\begin{figure}
\includegraphics[angle=0.,width=\hsize]{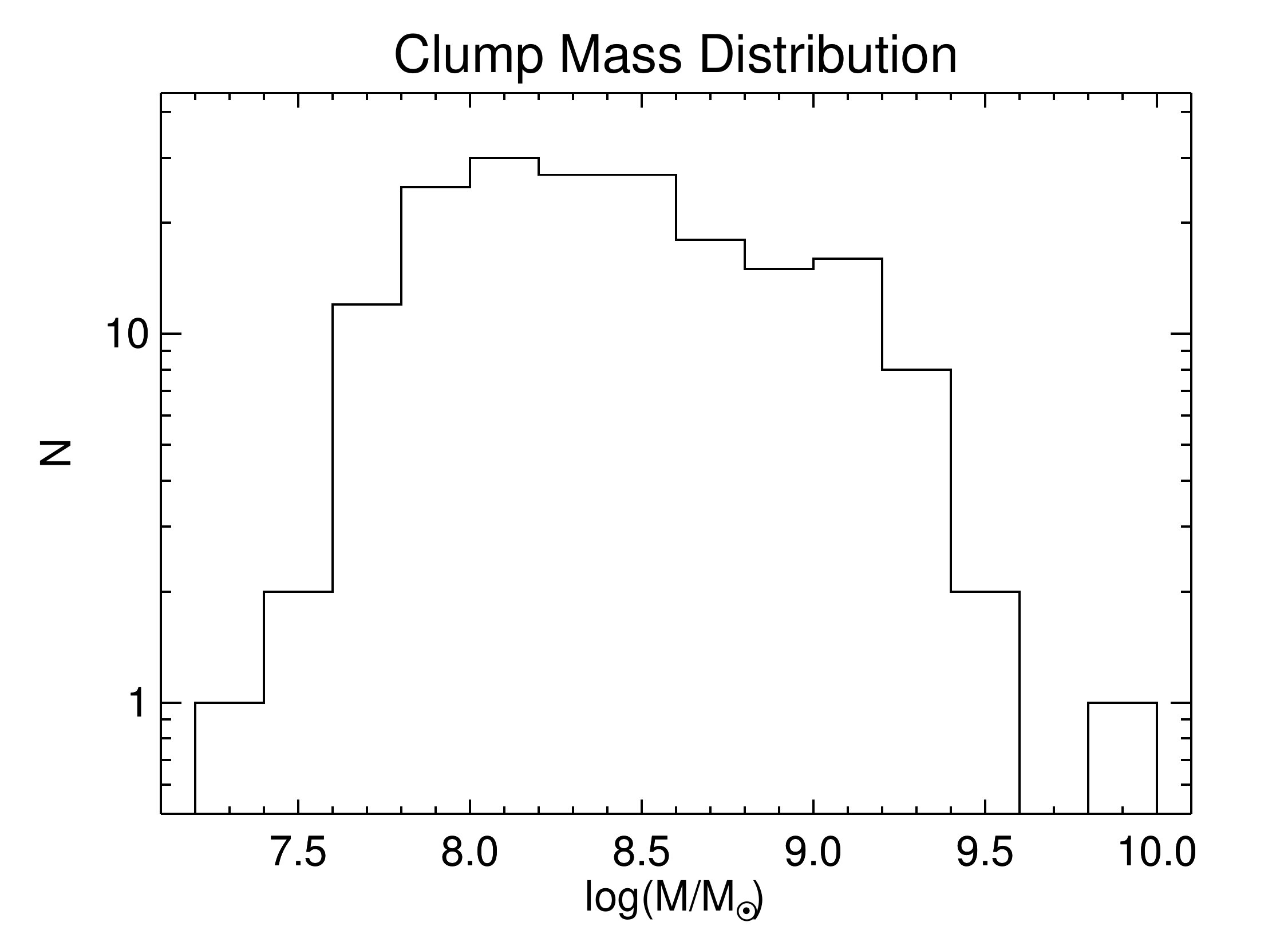}
\caption{The distribution of clump masses which ranges from $\sim$3$\times$10$^7$ to
10$^{10}$ $\Msun$.}
\label{f:clumpmassdist}
\end{figure}

The clump mass distribution is shown in Figure \ref{f:clumpmassdist}.
Clumps are detected in 100 Myr snapshots using a 10$\sigma$ threshold
in the 100 pc $\times$ 100 pc binned mass map with a 1.5 kpc smoothed
background removed.  The total clump mass is measured by summing up
the mass in a $500 \pc$ radius around the clump centre and removing
the background level determined in a $1.5-2.5 \kpc$ annulus.  Only
clumps with well-determined masses (error $\lesssim10\%$; to remove
noise fluctuations) and galactocentric radii greater than $100 \pc$
(to remove the inner bulge) are kept and these correspond well to the
visually identifiable clumps.  Note that because clumps survive for
$\sim500 \Myr$ before falling into the galactic centre, they are
detected multiple times in our technique.  The range of clump masses
is $3\times10^7$ to $8\times10^9\Msun$ with a mean of
$5\times10^8\Msun$, although our detection technique is not sensitive
to clumps below $\sim 3\times10^7$.  This mean mass is of the same
order, but higher than, the median clump mass found in high-redshift
galaxies observed at high resolution through gravitational lensing,
$10^8 \Msun$, \citep[e.g.][their Figure 2]{Cava2018}.

\subsection{A high redshift view}

Because our mass resolution results in only massive clumps being
identifiable in our simulation, we now explore whether our simulation
is an excessively clumpy galaxy when observed at the $\sim
\kpc$ resolution typical of galaxies at high redshift.  We use the 
ray-tracing 3D dust radiative transfer code DART-Ray
\citep{Natale2014, Natale2015, Natale2017}, which self-consistently 
calculates the effect of dust attenuation on the stellar radiation
(direct stellar light), and the subsequent dust emission (dust
re-radiated stellar light).  We use DART-Ray to produce face-on dust
attenuated images at rest UV, B and V bands of our simulated galaxy at
$t=1\Gyr$, shortly after the peak SFR in the high-\al\ sequence (see
Figure \ref{f:sfh}), as though it were a galaxy at $z = 2$.  The
direct light is assumed to be strictly due to star particles, whereas
emission by gas is ignored.  The stellar emissivity is computed at
wavelengths ranging from $0.1\mu$m to $30\mu$m for the mass, position,
age and metallicity of stars in the simulation at $1\Gyr$.  We adopt
spectral energy distributions from the {\sc starburst99} spectral
synthesis code assuming a Kroupa \citep{Kroupa2001} initial mass
function and the Padova AGB tracks.  For the optical properties of
grains and grain sizes we use the {\sc trust} benchmark dust model,
which is based on the model of \citet{Zubko2004}.
We assume that the gas \feh\ traces the metal abundance $Z$ according
to $Z = Z_{MW}10^{\feh}$ where $Z_{MW}$ is the metal abundance of the
Milky Way, assumed to be equal to 0.018.  We assumed that the dust
cross-section $C^{gas}_\lambda$ is proportional to the metal abundance
through a linear relation:
\begin{equation}
C^{gas}_\lambda = C^{gas}_{\lambda,MW} \frac{Z}{Z_{MW}}
\end{equation}
where $C^{gas}_{\lambda,MW}$ is the dust cross-section per unit gas
mass derived from the {\sc trust} dust model.
Then the radiation transfer is computed, reaching a highest spatial
resolution of $82 \pc$.

The images from DART-Ray are noise-free and have a spatial resolution
comparable to that of the simulation. To enable a better comparison
with real observations, we degrade the signal-to-noise and spatial
resolution of the images to that typical of high-redshift surveys
\citep[see also][]{Snyder2015}. Therefore, we generate synthetic {\it
Hubble Space Telescope} Wide Field Camera 3 ({\it HST}/WFC3) images
matched to the observing conditions of the CANDELS \citep{CANDELS2011}
and HUDF09 \citep{Bouwens2010} surveys, adopting the typical depth and
resolution of these surveys as listed in Table 1 of
\citet{Guo2013}. We create synthetic {\it HST}/WFC3 images for three
WFC3 filters --- F098M, F125W and F160W. At $z=2$, these trace 0.3 -
0.5 $\mu$m in the rest-frame, i.e., UV-, B- and V-band.
To generate the synthetic images, we first rescale the surface
brightness of each pixel as $I_{observed} =
I_{emitted}\times(1+z)^{-4}$ to account for cosmological surface
brightness dimming, ignoring k-corrections. The angular sizes of the
pixels are then scaled with the redshift and the pixels are rebinned
to 0.\arcsec6/pixel --- the typical pixel scale of drizzled {\it HST}
WFC3 observations. Next, we convolve the images with a 2D Gaussian
kernel to simulate the WFC3 point spread function, FWHM 0.\arcsec13
for F098M and 0.\arcsec16 for F125W and F160W. Finally, we add shot
noise to each pixel to match the typical depth of each survey ---
5$\sigma$ limits of $\sim$ 27.5 and 28.3 mag/arcsec$^2$ for the wide
and deep portions of the CANDELS survey, respectively, and 29.7
mag/arcsec$^2$ for the HUDF09.

\begin{figure*}
\includegraphics[angle=0.,width=\hsize]{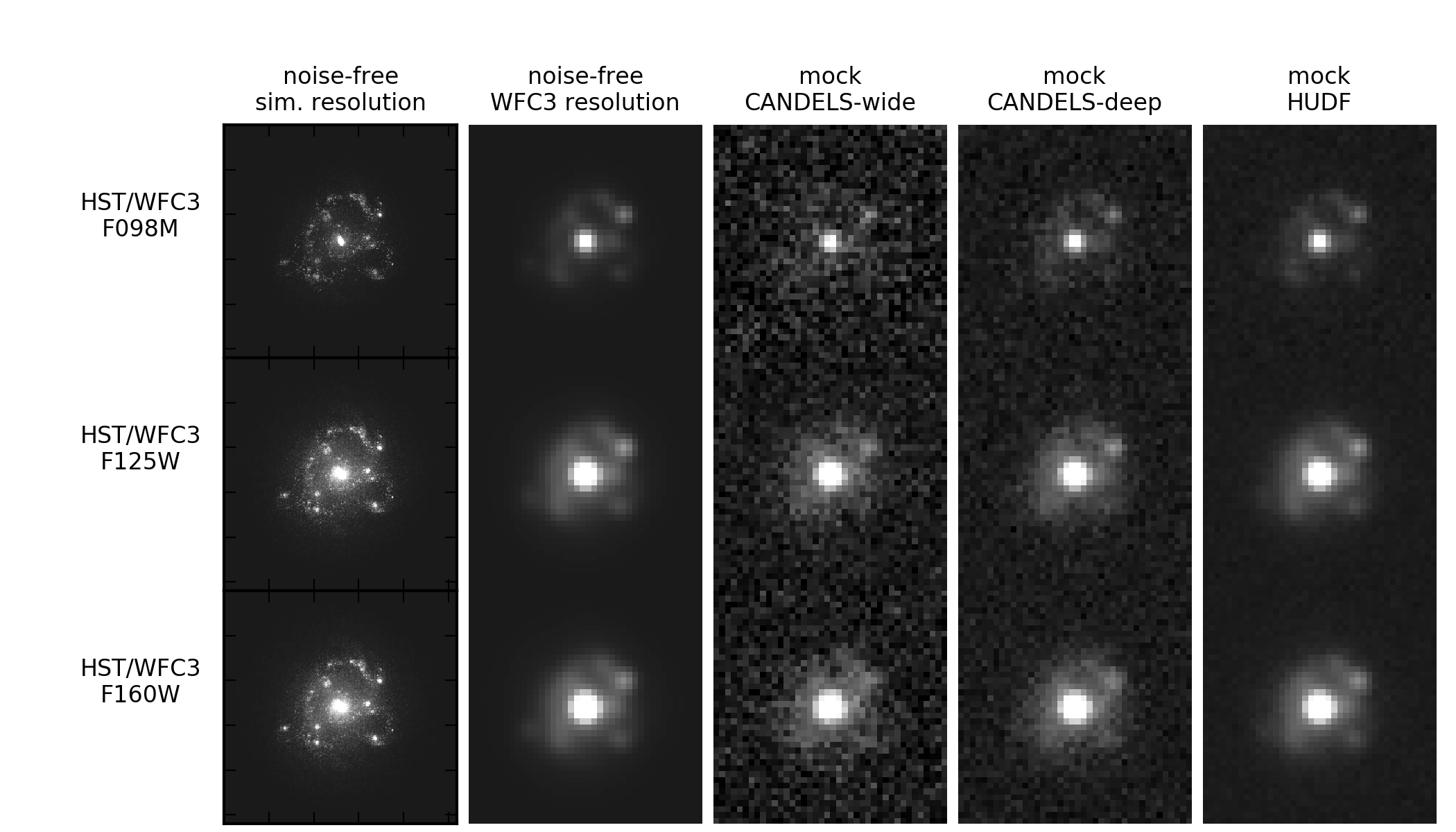}
\caption{The model at $t = 1 \Gyr$ as it would appear at $z=2$ if viewed by 
{\it HST}.  The left column shows the full high resolution images
produced by the radiative transfer calculations, while the remaining
columns convolve these images to the WFC3 resolution and add noise as
in CANDELS and HUDF.}
\label{f:hst}
\end{figure*}

Figure \ref{f:hst} shows the resulting images, at the original
resolution, and convolved with {\it HST} resolution, with surface
brightness dimming included and with noise added as in the CANDELS
survey \citep{CANDELS2011} and the Hubble Ultra Deep Field
\citep[HUDF,][]{HUDF2006}.
We use these images to identify clumps.  We first fit a
single-component S\'ersic \citep{Sersic1968} model to the F098M image
(0.3 $\mu$m rest wavelength) and search in the residual image for
off-centre clumps.  Following the definition of \citet{Guo2018}, we
identify one clump containing $\sim 14\%$ of the total UV flux in the
CANDELS-wide image.  This clump, visible in the top right of the
galaxy, is comprised of 4 sub-clumps which merge together at this
resolution.  It is at $4 \kpc$ from the galaxy centre at this time and
has a F098M-F160W ($U-V$ rest wavelength) colour of $0.34 \pm 0.15$,
which is significantly bluer than the rest of the galaxy, but within
the range of observed clumps \citep[e.g.][]{Guo2018}.
Our analysis of the mock deep and ultra-deep images identifies three
additional mini-clumps, having less than $1\%$ of the total UV flux
and would not be considered clumps in the usual definition.  Their
$U-V$ colours range from $0.2$ to $0.8$.

We have repeated this analysis at $t=2\Gyr$ and at $t=3\Gyr$ (still
pretending that the galaxy is at $z=2$).  At $t=2\Gyr$ we find 2
clumps, each contributing $\sim 10\%$ of the UV flux, while we find
only 1 clump and only at the UDF depth for $t=3\Gyr$, contributing
$\sim 9\%$ of the UV flux.

This analysis shows that our simulation is not plagued by an excess
clump formation and would not be viewed as unusual if found amongst
galaxies at $z \sim 2$.

 \section{Discussion and Conclusions}
\label{s:conclusion}

We have shown that the early, gas-rich phase of galaxy formation,
which supports two modes of star formation, distributed and clumpy, as
generally observed by {\it HST} at high redshift, naturally gives rise
to a bimodal distribution in chemical space, similar to what is seen
in the Milky Way.  The clumpy mode of star formation has a star
formation rate density about 100 times higher than the distributed
mode and leads to rapid self-enrichment with high \al-abundance,
whereas the distributed star formation is responsible for less
\al-enchanced stars. As the stellar mass of the galaxy builds up, the
gas fraction drops, leading to less star formation taking place in
clumps, which naturally shuts off most of the formation of high-\al\
stars (with star formation in the bulge being the exception).

While the vertical distribution of stars in Figure~\ref{f:zrms} and
Figure~\ref{f:vertprof} is reminiscent of the thick disk in the Milky
Way, the density distribution of high-\al\ stars flares, which means
there is not a direct one-to-one match between our model and the Milky
Way.  The high-\al\ vertical scale-height is consistent with Milky Way
observations at larger radii, but it is too small in the inner galaxy
to match the MW.  This inconsistency could be a result of the
idealised nature of our simulation, which is isolated and thus
excludes the effects of galaxy accretion that would dynamically heat
the disk (and increase the scale-height) further and is most important
at early times, when most of the high-\al\ stars are forming.  The
inclusion of these effects would bring our model into closer agreement
with the observations.  Moreover, the age of stars in chemical space
are qualitatively different from those in the Milky Way, although this
affects the low-\al\ sequence (the ``thin'' disc) more than the
high-\al\ one.  The main result we highlight here is the fact that the
current simulation is naturally able to produce the chemical
bimodality that is seen in the surveys but which, to date, has been
challenging to produce in simulations.

Clumps are thought to form via gravitational instability in gas rich
discs \citep{Noguchi1999, Immeli2004, Elmegreen2008, Inoue2016},
although some clumps may have an ``ex-situ'' (merger) origin
\citep[e.g.][]{Puech2009, Hopkins2013, Mandelker2014}.  The fate of 
clumps in simulations has been a subject of debate, with either their
destruction by supernova and/or radiative feedback \citep{Hopkins2012,
Genel2012, Buck2017, Oklopcic2017} or sinking to the centre of the
galaxy to contribute to the bulge \citep{Noguchi1999, Immeli2004,
Bournaud2007, Genzel2008, Elmegreen2008, Dekel2009, Ceverino2010,
Mandelker2017} suggested.  We have found that our clumps sink to the
centre of the galaxy on timescales of $200-500$ Myr.  In a companion
paper, we explore the consequences of clumps falling to the centre for
the chemistry of the bulge, which further helps constrain this model.

We have verified that if we increase the supernova feedback in our
simulations that clump formation is substantially inhibited and no
chemical bimodality results.  Although we use a lower feedback
efficiency than is often employed in cosmological simulations, we have
demonstrated that the clumps in our simulation are reasonable compared
to observations, such as those in the CANDELS-wide \citep{CANDELS2011}
and the HUDF \citep{HUDF2006}, and that our model is not excessively
clumpy.  At present there is some uncertainty regarding the overall
importance of clumps; our results suggest that even with modest clumps
the chemical consequences for the MW could be significant.

Recently \citet{Mackereth2018} suggested that a bimodality in chemical
space can be produced through gas accretion.  Their cosmological
simulations produce chemical bimodality less than half the time.  In
contrast, the scenario presented here will work for a large fraction of
galaxies with mass comparable to the Milky Way; \citet{Wuyts2012} find
that $\sim 40\%$ of galaxies of Milky Way mass are clumpy at $1.5 < z
< 2.5$ when considering their mass distribution (rising to $60\%$ in
the UV) in which case most such galaxies will have chemical
bimodalities.  At the moment this prediction cannot be tested but
future facilities, such as the Extremely Large Telescope, Giant
Magellan Telescope and the Thirty Meter Telescope, may be able to map
the chemical space of stars in nearby galaxies.

\subsection{Summary}

The results and implications of this study are as follows:
\begin{enumerate}

\item We produce a bimodal distribution of stars in the
  \ofe-\feh\ space via clumpy$+$distributed star formation.  The
  clumps self-enrich in $\alpha$-elements due to their high star
  formation rate density and produce the high-$\alpha$ sequence while
  the low-$\alpha$ sequence is produced by the distributed star
  formation.  The clumpy mode becomes less efficient as the gas mass
  fraction drops, leaving an \al-enhanced population that is old at
  the present day.  The clumps produced in our simulation are in
  reasonable agreement with clumps observed at $z \sim 2$
  \citep[e.g.][]{Guo2018}.

\item Clumps form on the low-\al\ sequence, and ascend to the 
high-\al\ sequence and their star formation rate increases.  Towards
the ends of their lives, they tend to drop back to the low-\al\
sequence, although this typically happens close to the bulge.

\item The two \al-sequences form simultaneously early on in the
  evolution of the Milky Way, resulting in overlapping ages.  Recent
  results by \cite{SilvaAguirre2017} and \citet{Hayden2017b} indicate
  that temporal overlap of the two sequences is seen in the
  observations as well.  For the foreseeable future this age overlap
  is the most promising way to test this hypothesis further.

\item The model predicts that bimodal chemical thick discs should be
  common at the mass of the Milky Way, since at least $\sim 40\%$ of
  galaxies of Milky Way mass are clumpy at $z \sim 2$
  \citep{Wuyts2012}.  An important test of this scenario therefore is
  that chemical bimodality should be common in Milky Way mass
  galaxies.
  
\end{enumerate}

The importance of clumps to the evolution of disc galaxies continues
to be hotly debated \citep[e.g.][]{Ceverino2010, Hopkins2012,
Moody2014, Tamburello2015, Mayer2016, Mandelker2017, Buck2017,
Tamburello2017, Benincasa2018}.  The lifetimes and masses of clumps
are therefore somewhat uncertain, and in simulations is dependent on
resolution and sub-grid physics \citep[e.g.][]{Tamburello2015}, and
possibly also on the mode of gas accretion \citep{Ceverino2010}.
However there is a growing view from high redshift observations that
clumps cannot be too massive given the high degree of rotation in
discs \citep[e.g.][]{Wuyts2012, Wisnioski2015, Wisnioski2018}.  For
instance, by seeding their models with a spectrum of velocity
perturbations matching the turbulent cascade, \citet{Benincasa2018}
found that likely bound masses cannot be larger than $10^9 \Msun$ and
typical masses $\sim 10^6\Msun$ to $\sim 10^8 \Msun$.  At lower mass
scales, simulations of giant molecular clouds have shown that
photoionizing radiation, stellar-winds, and supernova feedback reduce
star formation by only a modest amount \citep[e.g.][]{Rogers2013,
Dale2017, Howard2017}, indicating that understanding the role of
feedback in controlling clumps is still an open question.

The high resolution, the inclusion of metal-line cooling and the
absence of any ab initio stars make our simulation susceptible to
clump formation.  Our results in this and the companion papers show
that clumps may solve some important problems in galaxy formation that
have not been explained fully satisfactorily otherwise.  Nonetheless,
the model in this paper is quite simple and there is much room for
improvement.  In particular we hope that future works will provide
even more realistic models of clumps which will permit more detailed
comparisons with data, rather than the proof-of-concept approach we
have taken here.

 \acknowledgements
\label{s:ack}

V.P.D. is supported by STFC Consolidated grant \#~ST/R000786/1 and
acknowledges the personal support of George Lake during part of this
project, as well as the Pauli Center for Theoretical Studies, which is
supported by the Swiss National Science Foundation (SNF), the
University of Z\"urich, and ETH Z\"urich.
S.R.L. acknowledges support from the Michigan Society of Fellows.
S.R.L. was also supported by NASA through Hubble Fellowship grant
HST-HF2-51395.001-A from the Space Telescope Science Institute, which
is operated by the Association of Universities for Research in
Astronomy, Incorporated, under NASA contract NAS5-26555.
DBF acknowledges support from ARC Future Fellowship FT170100376.
V.P.D. anknowledges the importance of ideas developed during the
workshops ``Disk instabilities across cosmic scales'' (Sexten, July,
2017) and ``Thin, thick and dark disks'' (Ascona, July, 2017).
V.P.D. thanks Lucio Mayer for his support in attending the former, and
the Congressi Stefano Franscini for their financial support in
organising the latter.
The authors thank Giovanni Natale for his help with the use of
DART-Ray.
The simulation in this paper was run at the DiRAC Shared Memory
Processing system at the University of Cambridge, operated by the
COSMOS Project at the Department of Applied Mathematics and
Theoretical Physics on behalf of the STFC DiRAC HPC Facility
(www.dirac.ac.uk). This equipment was funded by BIS National
E-infrastructure capital grant ST/J005673/1, STFC capital grant
ST/H008586/1 and STFC DiRAC Operations grant ST/K00333X/1. DiRAC is
part of the National E-Infrastructure.  Images with DART-Ray were
built using the High Performance Computing Platform of Peking
University.
This work was completed at Aspen Center for Physics, which is
supported by National Science Foundation grant PHY-1607611. This work
was partially supported by a grant from the Simons Foundation.
We thank the anonymous referee for their useful comments that helped
improve this paper.


\bibliography{ref}
\bibliographystyle{aj}

\end{document}